\documentclass[letterpaper,aps,prd,twocolumn,tightenlines,preprintnumbers,nofootinbib,showkeys,superscriptaddress]{revtex4-1}
\pdfoutput=1

\usepackage{XCharter}
\usepackage{mathptmx}
\usepackage[T1]{fontenc}

\usepackage{fullpage}
\usepackage{amsfonts}
\usepackage{amsmath}
\usepackage{slashed}
\usepackage{amssymb}
\usepackage{graphicx}
\usepackage{epic}
\usepackage{eepic}
\usepackage{epsfig}
\usepackage{latexsym}
\usepackage[dvipsnames]{xcolor}
\usepackage[export]{adjustbox}
\usepackage{float}
\usepackage{multirow}
\usepackage{hyperref}
\usepackage{enumitem}
\usepackage{lipsum}
\hypersetup{colorlinks=true,citecolor=red,linkcolor=NavyBlue,urlcolor=NavyBlue}
\usepackage[caption=false,labelformat=simple]{subfig}
 
\usepackage{natbib}
\usepackage{relsize}
\usepackage[left=1.6cm,right=1.6cm,top=1.65cm,bottom=1.65cm]{geometry}
\usepackage{comment}
\usepackage{shorthand}
\usepackage{dcolumn}

\begin{document}

\title{Right-handed neutrino pair production via second-generation leptoquarks}

\author{Arvind Bhaskar}
\email{arvind.bhaskar@research.iiit.ac.in}
\affiliation{Center for Computational Natural Sciences and Bioinformatics, International Institute of Information Technology, Hyderabad 500 032, India}

\author{Yash Chaurasia}
\email{yash.chaurasia@research.iiit.ac.in}
\affiliation{Center for Computational Natural Sciences and Bioinformatics, International Institute of Information Technology, Hyderabad 500 032, India}

\author{Kuldeep Deka}
\email{kuldeepdeka.physics@gmail.com}
\affiliation{Department of Physics and Astrophysics, University of Delhi, Delhi 110 007, India}

\author{Tanumoy Mandal}
\email{tanumoy@iisertvm.ac.in}
\affiliation{Indian Institute of Science Education and Research Thiruvananthapuram, Vithura, Kerala, 695 551, India}

\author{Subhadip Mitra}
\email{subhadip.mitra@iiit.ac.in}
\affiliation{Center for Computational Natural Sciences and Bioinformatics, International Institute of Information Technology, Hyderabad 500 032, India}

\author{Ananya Mukherjee}
\email{ananyatezpur@gmail.com}
\affiliation{Theory Division, Saha Institute of Nuclear Physics, 1/AF Bidhannagar, Kolkata
700 064, India} 

\begin{abstract}\noindent
No direct experimental constraints exist on Leptoquark (LQ) couplings with quarks and right-handed neutrinos (RHNs). If a LQ dominantly couples to RHNs, it can leave unique signatures at the LHC. The RHNs can be produced copiously from LQ decays as long as they are lighter than the LQs. LQ-induced RHN production has never been searched for in experiments. This channel can act as a simultaneous probe for RHNs and LQs that dominantly couple to RHNs. In this paper, we consider all possible charge-$2/3$ and $1/3$ scalar and vector LQs that dominantly couple to second-generation quarks and RHN. We study the pair and single productions of TeV-scale LQs and their subsequent decay to sub-TeV RHNs, realised in the inverse seesaw framework. We also consider RHN pair production through a $t$-channel LQ exchange. The single LQ production and $t$-channel contributions can be significant for large LQ-RHN-quark couplings. We systematically combine events from these processes leading to a pair of RHNs plus jets to study the prospects of LQ-assisted RHN pair production. We analyse the monolepton and opposite-sign dilepton final states and estimate the discovery reach at the high-luminosity LHC.
\end{abstract}
\maketitle 

\section{Introduction}
\relscale{0.97}
\label{sec:intro}
\noindent
The neutrino oscillation data provide vital evidence for the presence of physics beyond the Standard Model (SM) by pointing towards the neutrino masses. However, probing the neutrino sector at particle colliders is not simple---due to their elusive nature, the light neutrinos pass through the detectors undetected. Generally, one considers heavy right-handed neutrinos (RHNs, $\n_R$'s) to generate light-neutrino masses through the seesaw mechanism. Observing RHNs at the Large Hadron Collider (LHC) would be an important milestone since it could shed light on the neutrino mass generation mechanism and various serious issues like the dark matter problem, matter-antimatter asymmetry, etc. In principle, the LHC could probe RHNs if they are within the TeV range. However, a straightforward application of the vanilla seesaw mechanism (Type-I)~\cite{Minkowski:1977sc,Mohapatra:1979ia} puts the RHNs several orders of magnitude above the TeV scale. 

There are other mechanisms to generate the masses---like the inverse seesaw mechanism (ISM)~\cite{Mohapatra:1986aw,Mohapatra:1986bd}---where the RHNs can be within the reach of the LHC. But, even the TeV-scale RHNs are not easy to produce at the LHC (see Ref.~\cite{Banerjee:2015gca} for the prospect study of RHNs at $e^+e^-$ colliders). Since they are singlet under the SM gauge group, they interact feebly with the SM fields through their small overlaps with the left-handed neutrinos generated after the electroweak symmetry breaking. RHNs can be produced through the decay of another particle like $W^\prime$~\cite{Keung:1983uu,ThomasArun:2021rwf}, $Z^\prime$~\cite{Ekstedt:2016wyi,Choudhury:2020cpm,Deka:2021koh,Arun:2022ecj}, etc. Since the LHC is a hadron collider, RHNs can also be produced through an intermediary that connects with the strong sector. In this paper, we consider one that fits naturally in this picture, namely, the leptoquark (LQ). 

LQs are hypothetical coloured scalar or vector bosons with both baryon and lepton numbers. Hence, they can act as connectors between the baryon and lepton sectors. They appear in various beyond-the-SM theories like the Pati-Salam models~\cite{Pati:1974yy}, Grand Unified Theories~\cite{Georgi:1974sy}, various compositeness theories~\cite{Schrempp:1984nj}, or $R$-parity-violating Supersymmetry~\cite{Barbier:2004ez}, etc. Nowadays, they are popular in the literature for explaining various experimental anomalies, like the ones seen in $R_{D^{(*)}}$,  $(g-2)_\mu$, $W$-mass, etc.~\cite{Calibbi:2017qbu,Blanke:2018sro,Azatov:2018kzb,Mandal:2018kau,Aydemir:2019ynb,Balaji:2019kwe,Crivellin:2019dwb,Crivellin:2020ukd,Crivellin:2020tsz,Bhaskar:2021pml,Bhaskar:2022vgk}. The phenomenology of LQs and their discovery prospects at different colliders can be found in Refs.~\cite{Das:2012ze,Dumont:2016xpj,Das:2017nvm,Dey:2017ede,Bandyopadhyay:2018syt,Das:2018usr,Chandak:2019iwj,Padhan:2019dcp,Bhaskar:2020gkk,Iguro:2020keo,Ghosh:2022vpb,Desai:2023jxh}.

The LHC has an ongoing LQ-search program. Both ATLAS and CMS Collaborations are looking for single and pair productions of LQs in various combinations of leptons ($\ell$ or $\n_L$) and jets (light or $t$ jet) in the final states. The current mass-exclusion limits on scalar and vector LQs go up to $1.73$~\cite{ATLAS:2020dsk} and $1.98$~\cite{ATLAS:2022wcu} TeV, respectively. There are also indirect bounds on LQ couplings (LQ-quark-lepton) from the high-$p_T$ dilepton or monolepton with missing energy data~\cite{Mandal:2018kau,Bhaskar:2021pml}. A LQ can simultaneously couple with a SM quark and a RHN. However, none of the direct or indirect bounds concerns this coupling. A LQ can decay to a RHN+jet final state if the RHN is lighter than the LQ. If the LQ decays exclusively (or predominantly) through this mode, it becomes unrestricted by all the direct or indirect bounds. In other words, a large part of the LQ parameter space remains unexplored and unrestricted by the LHC. 

In this paper, we utilise this freedom to study the production of RHNs via LQs at the hadron collider (similar or related phenomenological studies are found in Refs.~\cite{Das:2017kkm,Mandal:2018qpg,Bhaskar:2020kdr,Cottin:2021tfo,Bhaskar:2022ygp}). If we assume the LQ couples with no other leptons except the RHNs, there are two possible production processes for the RHNs---they can come from LQ decays, or they can be directly pair produced by $t$-channel LQ exchanges in the quark fusion mode. As we discuss later, normally, the LQ-decay mode is more promising than the $t$-channel LQ exchange if the RHNs are lighter than the LQ. One reason is that LQs can be produced copiously through strong interaction. Secondly, the cross-section of the $t$-channel LQ-exchange process is susceptible to the LQ-quark-lepton coupling (goes as the fourth power); unless the coupling is of order one or larger, its cross-section is relatively smaller. Hence, in this paper, we mainly focus on RHN production through LQ decay. 

For our purpose, we consider a setup where the light neutrinos get their masses through the ISM. In the ISM, there are three RHNs (one for each generation) and three extra neutral fermions ($S_{L_i}$ with $i\in\{1,2,3\}$ being the generation index, all singlets under the SM gauge group). Due to these extra fermions, the RHNs can have (sub-)TeV-range masses. We are interested in the parameter regions where the RHNs are lighter than LQs so they can decay exclusively through the RHN+jet decay mode. For the LHC to detect their signatures, the RHNs should not be long-lived and decay to SM particles within the detectors. Generally, the strongest collider bounds on the RHNs come from the searches for the same-sign dilepton pairs, the signature of the Majorana nature of the RHNs. However, these bounds do not affect our analysis as the RHNs are pseudo-Dirac types in ISM. They decay mainly through the $\nu_R \rightarrow W^{\pm}\ell^{\mp}$ and $\nu_R \rightarrow Z/h~\nu_{\ell}$ processes in roughly $2:1:1$ ratio.

Generally, LQs can have inter or intra-generational couplings (i.e., the quark and the lepton that couple to a LQ need not be of the same generation). LQs that dominantly couple to third-generation fermions should be separately searched for from those that mostly couple to the lighter-generation fermions~\cite{Chandak:2019iwj,Bhaskar:2020gkk,Bhaskar:2021gsy}. This is because the detection strategies for the third-generation fermions differ significantly from the first two generations. There are also significant differences in the single and indirect LQ-production (i.e., $t$-channel LQ exchange) cross-sections depending on whether they couple to the first or second-generation quarks. This happens because the first-generation quarks have the largest parton distribution functions (PDFs). For the LHC analysis, we mainly consider second-generation interactions, i.e., LQs essentially decay to a second-generation quark and a second-generation RHN (i.e., second-generation LQs), and the RHN decays to produce a second-generation lepton in the final state. There are two reasons for this choice. First, the muon-detection efficiency is better than the electron detection at the LHC. Second, and more importantly, the second-generation case gives us a conservative estimate of the prospects of this channel. Due to the larger PDFs of the up and down quarks, the LHC discovery reach for mostly first-generation interactions would be better.  We shall report the prospects for the first and third-generation cases separately.

The paper is organised as follows. We review the ISM in the next section. In Sec.~\ref{sec:lqmodel}, we list the possible LQ models with RHN-decay mode and introduce some simple phenomenological models. In Sec.~\ref{sec:lhc}, we discuss the signals and the backgrounds and present our results in Sec.~\ref{sec:HL-LHC}. Finally, we conclude in Sec.~\ref{sec:conclu}.

\section{The Inverse Seesaw Mechanism}
\noindent We can write the neutrino-sector interactions as
\begin{equation}\label{ISS_Lag}
-\mathcal{L} \supset \lambda_{\nu}^{  i} \,\overline{L_i} \,\widetilde{H} \,\n_{R_i} + M_{R} \,\overline{\n_{R_i}} \,S_{L_i}+ \frac{1}{2} \mu\, \overline{S_{L_i}^c} S_{L_i} +\textrm{H.c.},
\end{equation}
where $i$ is the generation index, $L_i$ is the $i$th lepton doublet, $\widetilde{H} = i \sigma_2 H^*$, and the superscript $c$ denotes charge conjugation. Sterile neutrinos are denoted as $S_L$. They are SM singlets and carry the lepton number $L=1$. Similarly, RHNs also have $L=1$. The $S_L$ fields interact with $\nu_R$ but not directly with the SM leptons. However, tiny interactions can arise after mass mixing. In the $\{ {\mathbf\nu}_L^{c}\ {\mathbf\n}_R\  {\mathbf S}_L^{c}\}$ basis, the neutrino mass matrix  can be written as
\begin{equation}
{\mathbf M_\nu} = \begin{pmatrix}\label{ISS_matrix}
{\mathbf 0} && {\mathbf m_D} && {\mathbf 0} \\
{\mathbf m_D^T} && {\mathbf 0} && {\mathbf M_R} \\
{\mathbf 0} && {\mathbf M_R^T} && {\mathbf \mu} \\
\end{pmatrix},
\end{equation}  
where all of ${\mathbf m_D},\, {\mathbf M_R}\,\, \text{and}\,\, {\mathbf \mu}$ are $3\times3$ mass matrices. The light-neutrino masses are obtained after block diagonalising the mass matrix as the following,
\begin{equation}\label{eq:mnu}
{\mathbf m_\nu} \approx {\mathbf m_D ({\mathbf M_R^T})^{-1}{\mathbf \mu} {\mathbf M_R^{-1}}{\mathbf m_D^T}}.
\end{equation}
For the heavy components, the $6\times6$ mass matrix in the $({\mathbf N_R}\ {\mathbf S})$ basis can be written as
\begin{equation}
{\mathbf M_\nu^{6\times6}} = \begin{pmatrix}
{\mathbf 0} && {\mathbf M_R} \\
{\mathbf M_R^T} && {\mathbf \mu}  \\
\end{pmatrix},
\end{equation}
where $\mu\sim\textrm{keV}$ is the lepton-number violating scale that essentially acts as the source of tiny non-degeneracy among the final pseudo-Dirac pairs. 

For us, the essential point is our collider study is largely insensitive to the parameters in the neutrino sector (like ${\mathbf m_D}$, ${\mathbf M_R}$, ${\mathbf \m}$ etc.) except the mass and decays of the second-generation RHN. Hence, we do not need any specially tuned parameter as sub-TeV RHNs decaying to $W^{\pm}\ell^{\mp}$ and $Z/h~\nu_{\ell}$ final states are easily found in the allowed parameter space (see, e.g.,~\cite{ThomasArun:2021rwf,Arun:2022ecj}).

\section{scalar and Vector Leptoquark Models}\label{sec:lqmodel}
\noindent
We list the LQs---scalars (sLQ) and vectors (vLQs)---with interactions with the RHNs~\cite{Dorsner:2016wpm}. We ignore the diquark operators to bypass the proton-decay constraints. 

\subsection{Scalar LQs}
\label{sec:scalarmodels}
\noindent
$\blacksquare\quad$\underline{$\tilde{R}_2=(\overline{\mathbf{3}},\mathbf{2},1/6)$:}
The interaction of $\tilde{R}_2$ can be written as follows,
\begin{align}
\label{eq:LagR2}
\mathcal{L} \supset &~ \tilde{y}^{\overline{LR}}_{2\,ij}~\bar{Q}_{L}^{i,a} \tilde{R}_{2}^{a} \nu_{R}^{j}+\textrm{H.c.},
\end{align}
where $\bar{Q}_L$ denotes the left-handed quark doublet, $a,b=1,2$ are the $SU(2)$ indices, and $\epsilon = i\sigma^2$. The terms relevant to our analysis are
\begin{align}
\mathcal{L} \supset &\ \tilde{y}^{\overline{LR}}_{2  \ ii}~\bar{u}_{L}^i \nu_{R}^i \tilde{R}_2^{2/3} + \tilde{y}^{\overline{LR}}_{2  \ ii} ~\bar{d}_{L}^i \nu_{R}^i \tilde{R}_2^{-1/3} +\textrm{H.c.}\label{eq:LagR2sim}
\end{align}

\noindent $\blacksquare\quad$\underline{${S}_1=(\overline{\mathbf{3}},\mathbf{1},1/3)$:}
The only relevant term in the interaction Lagrangian of $S_1$ is
\begin{align}
\label{eq:LagS1}
\mathcal{L} \supset &~ -\bar{y}^{RR}_{1\,ii}~\bar{d}_{R}^{C~i} {S}_{1} \nu_{R}^{i}+\textrm{H.c.}
\end{align}

\noindent $\blacksquare\quad$\underline{${\overline{S}}_1=(\overline{\mathbf{3}},\mathbf{1},-2/3)$:}
The relevant term is
\begin{align}
\label{eq:LagS1bar}
\mathcal{L} \supset &~ +\bar{y}^{\overline{RR}}_{1\,ii}~\bar{u}_{R}^{C~i} {\Bar{S}}_{1} \nu_{R}^{i}+\textrm{H.c.}
\end{align}

\subsection{Vector LQs}
\label{sec:vectormodels}
\noindent
$\blacksquare\quad$\underline{$\tilde{V}_2=(\overline{\mathbf{3}},\mathbf{2},-1/6)$:}
The RHN interaction of $\tilde{V}_2$ can be written as
\begin{align}
\label{eq:LagV2}
\mathcal{L} \supset &~ \tilde{x}^{\overline{LR}}_{2\,ij}\bar{Q}_{L}^{C~i,a}\gamma^{\mu}\epsilon^{ab} \tilde{V}_{2,\mu}^{b} \nu_{R}^{j}+\textrm{H.c.},
\end{align}
which gives us the terms relevant to our analysis:
\begin{align}
\mathcal{L} \supset &\ \tilde{x}^{\overline{LR}}_{2  \ ii}~\bar{u}_{L}^{C~i}\gamma^{\mu} \nu_{R}^i \tilde{V}_{2,\mu}^{-2/3} - ~\tilde{x}^{\overline{LR}}_{2  \ ii} ~\bar{d}_{L}^{C~i}\gamma^{\mu} \nu_{R}^i \tilde{V}_{2,\mu}^{-1/3} +\textrm{H.c.}\label{eq:LagV2sim}
\end{align}

\noindent$\blacksquare\quad$\underline{${\bar{U}}_1=(\overline{\mathbf{3}},\mathbf{1},-1/3)$:}
The only relevant term for $\bar{U}_1$ is as follows,
\begin{align}
\label{eq:LagU1bar}
\mathcal{L} \supset &~ \bar{x}^{\overline{RR}}_{1\,ii}~\bar{d}_{R}^{i}\gamma^{\mu} {\bar{U}}_{1,\mu} \nu_{R}^{i}+\textrm{H.c.}
\end{align}

\noindent$\blacksquare\quad$\underline{${U}_1=(\overline{\mathbf{3}},\mathbf{1},2/3)$:}
The relevant term for ${U}_1$ is as follows,
\begin{align}
\label{eq:LagU1}
\mathcal{L} \supset &~ \bar{x}^{\overline{RR}}_{1\,ii}~\bar{u}_{R}^{i}\gamma^{\mu} {\bar{U}}_{1,\mu} \nu_{R}^{i}+\textrm{H.c.}
\end{align}

\subsection{Simple models}
\label{subsec:simplifiedmodels}
\noindent We can generically express the LQ interactions in terms of some phenomenological Lagrangians in the spirit of Refs.~\cite{Chandak:2019iwj, Bhaskar:2020gkk, Bhaskar:2021gsy} as
\begin{align} 
\label{eq:simpleSlag1}
\mathcal{L} \supset&\ \lambda_1 \bar{d}_{L}\nu_{R}\phi_1 + \lambda_2 \bar{u}_{L}\nu_{R}\phi_2 +{\rm H.c.}, \\
\label{eq:simplevlag1}  
\mathcal{L} \supset&\  
\Lambda_1\bar{d}_{R}\lt(\gamma\cdot\chi_1\rt)\nu_{R} + \Lambda_2\bar{u}_{R}\lt(\gamma\cdot\chi_2\rt)\nu_{R}+{\rm H.c.},
\end{align}
where $d$ and $u$ represent generic down and up-type quarks, respectively and $\phi_n$, $\chi_n$ denote an (absolute) charge-$n/3$ LQ.
If we assume no mixing among the right-handed quarks, the interactions of the weak-singlet LQs ($S_1$, $\bar S_1$, $U_1$, or $\bar U_1$) can be modelled by considering only one nonzero $\lambda_n$ or $\Lambda_n$ (flavour diagonal)---we assume these couplings are real for simplicity. However, to model the doublet ($\tilde R_2$ or $\tilde V_2$) interactions, we have to set $\lm_1 = \lm_2$ or $\Lm_1=-\Lm_2$ and, depending on whether the LQ interaction is aligned with the up-type or down-type quarks, replace $d^i_L$ (where $i$ indicates the generation) by $[V_{\rm CKM}]_{ij}d^j_L$ or $u^i_L$ by $[V_{\rm CKM}^{\dag}]_{ij}u^j_L$ where $V_{\rm CKM}$ is the Cabibbo-Kobayashi-Maskawa (CKM) quark-mixing matrix.
The kinetic terms of the vLQ Lagrangians contain an extra parameter, $\kappa$~\cite{Dorsner:2016wpm},
\begin{align}
\mc L \supset -\frac12\chi^\dag_{\m\n}\chi^{\m\n} + M^2_\chi\ \chi^\dag_\m\chi^\m -ig_s\kp \ \chi^\dag_\m T^a\chi_\n\ G^{a\,\m\n},\label{eq:lkin}
\end{align}
where $\chi_{\m\n}$ stands for the field-strength tensor of $\chi$. The cross-section of the pair and single production of vLQs depends on $\kappa$.

\begin{figure*}
\centering
\captionsetup[subfigure]{labelformat=empty}
\subfloat[\quad\quad\quad(a)]{\includegraphics[width=0.25\textwidth]{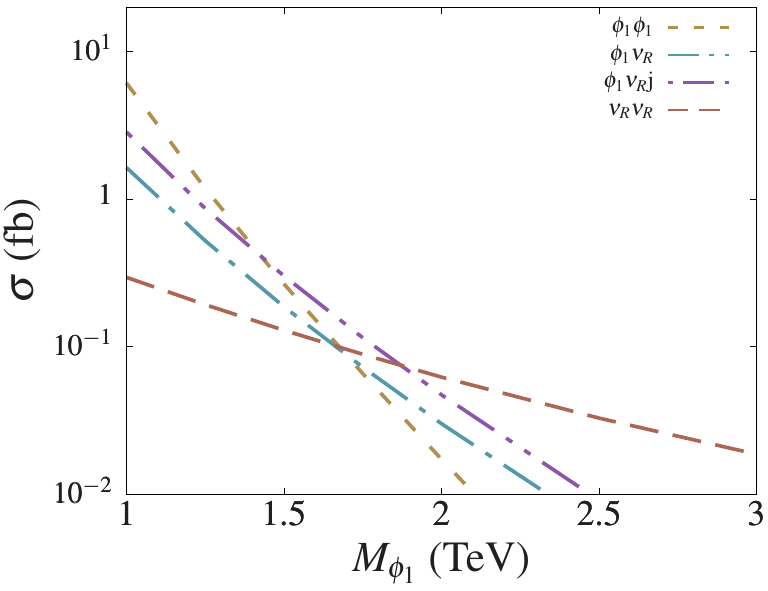}\label{fig:CSset101}}\hfill
\subfloat[\quad\quad\quad(b)]{\includegraphics[width=0.25\textwidth]{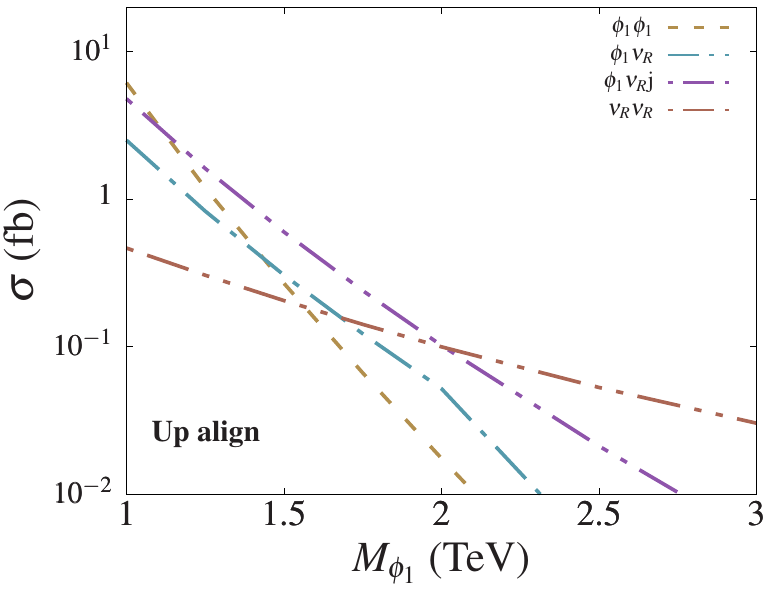}\label{fig:CSset102}}\hfill
\subfloat[\quad\quad\quad(c)]{\includegraphics[width=0.25\textwidth]{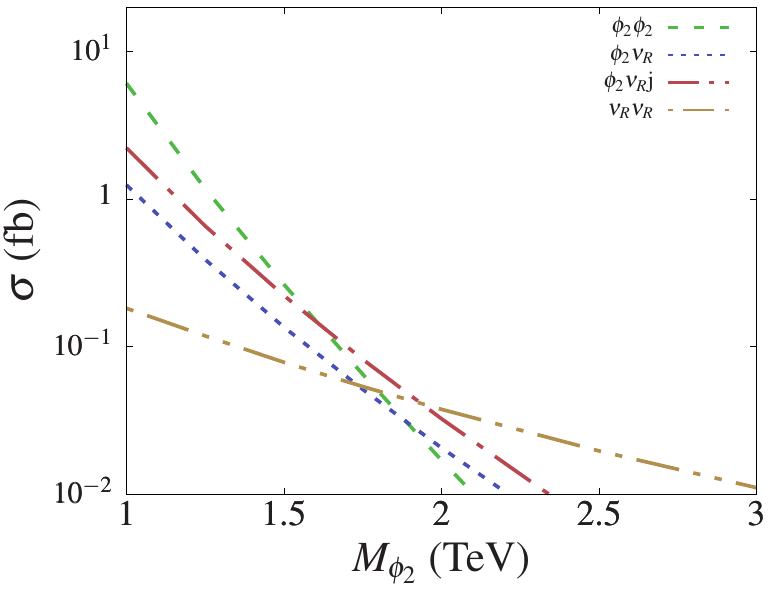}\label{fig:CSset103}}\hfill
\subfloat[\quad\quad\quad(d)]{\includegraphics[width=0.25\textwidth]{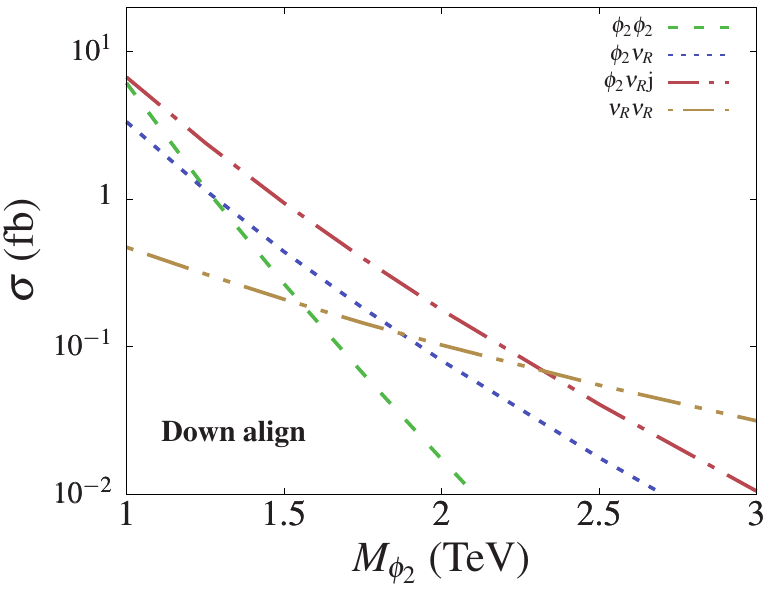}\label{fig:CSset104}}\\


\subfloat[\quad\quad\quad(e)]{\includegraphics[width=0.25\textwidth]{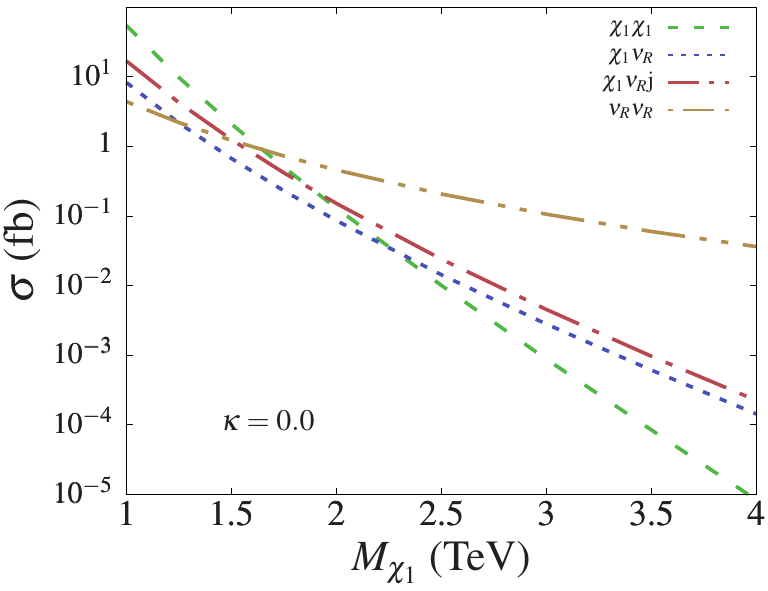}\label{fig:CSset201}}\hfill
\subfloat[\quad\quad\quad(f)]{\includegraphics[width=0.25\textwidth]{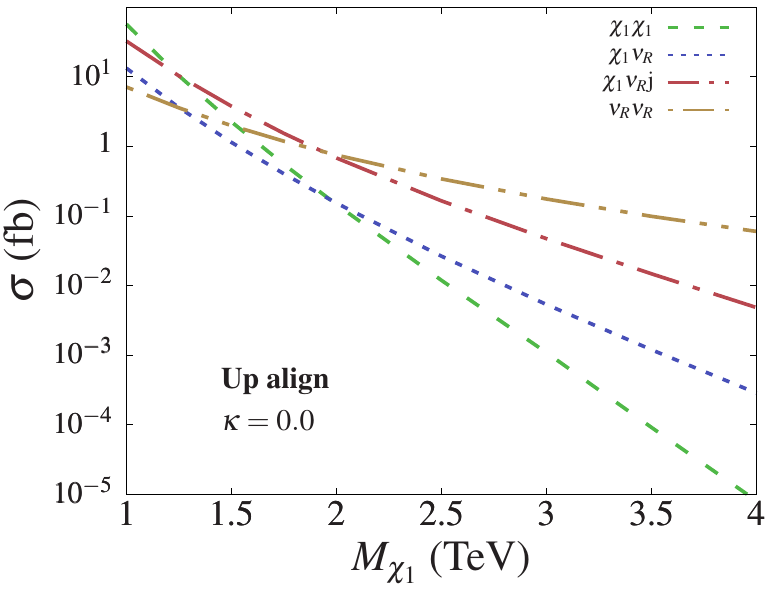}\label{fig:CSset202}}\hfill
\subfloat[\quad\quad\quad(g)]{\includegraphics[width=0.25\textwidth]{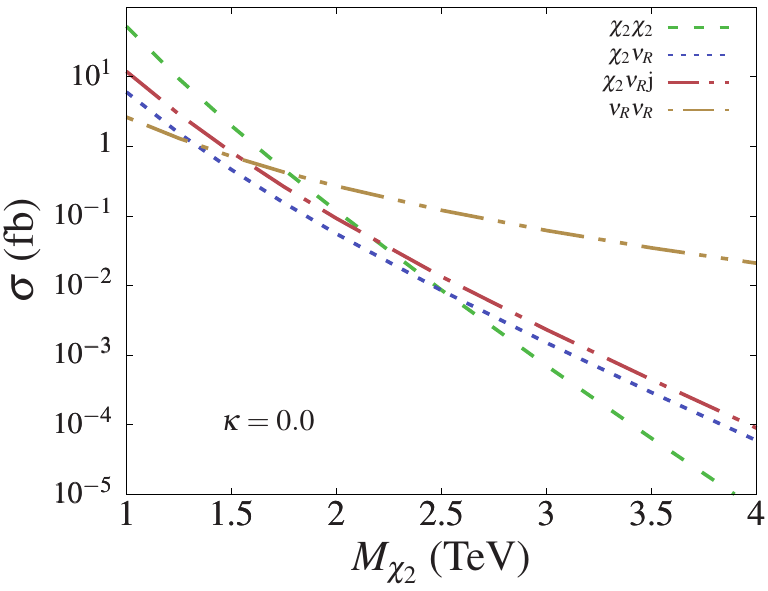}\label{fig:CSset203}}\hfill
\subfloat[\quad\quad\quad(h)]{\includegraphics[width=0.25\textwidth]{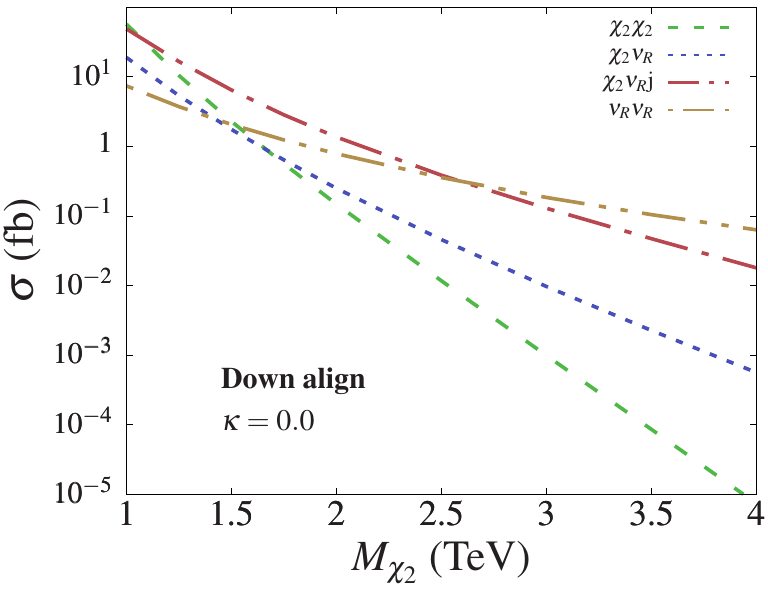}\label{fig:CSset204}}\\
\subfloat[\quad\quad\quad(i)]{\includegraphics[width=0.25\textwidth]{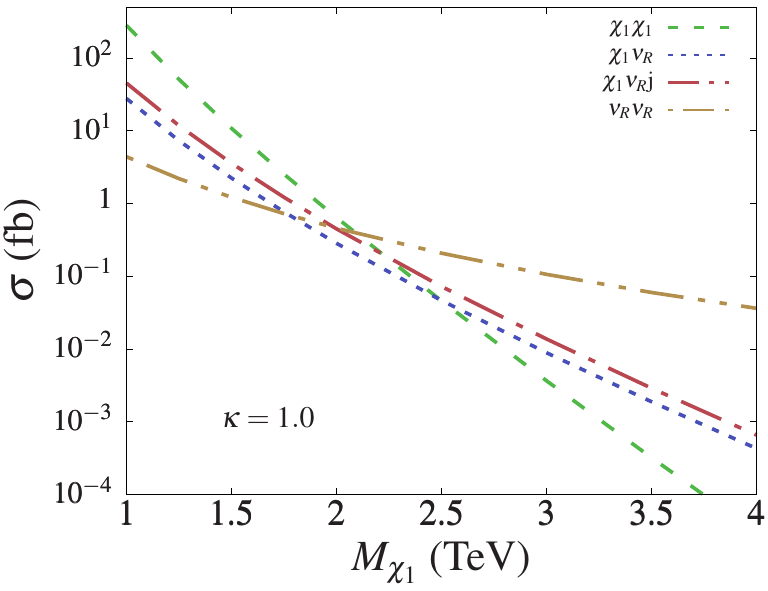}\label{fig:CSset301}}\hfill
\subfloat[\quad\quad\quad(j)]{\includegraphics[width=0.25\textwidth]{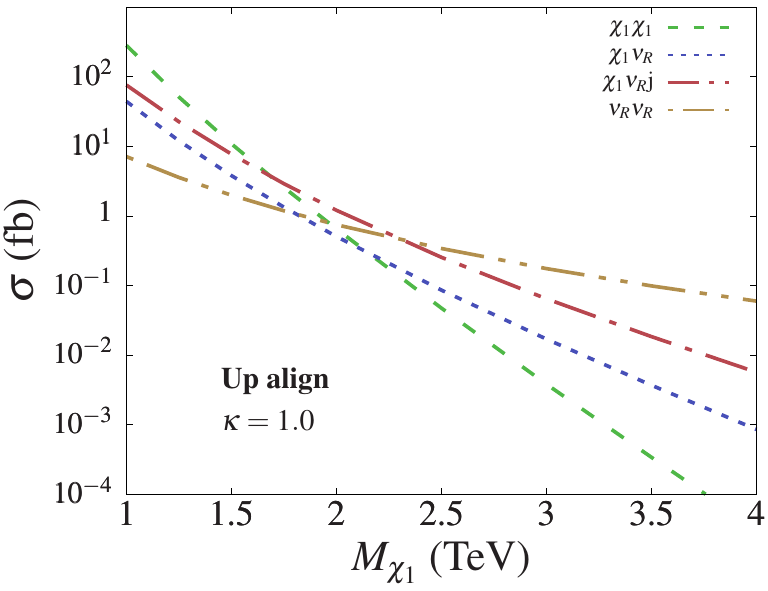}\label{fig:CSset302}}\hfill
\subfloat[\quad\quad\quad(k)]{\includegraphics[width=0.25\textwidth]{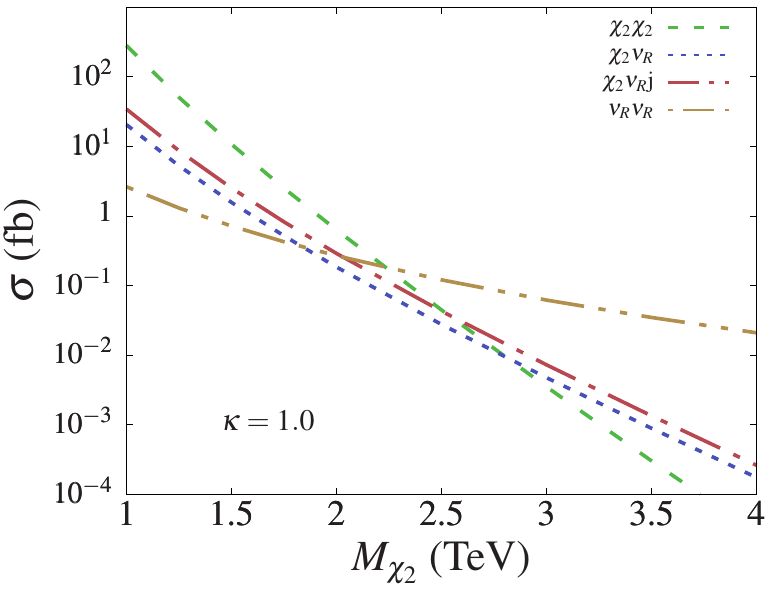}\label{fig:CSset303}}\hfill
\subfloat[\quad\quad\quad(l)]{\includegraphics[width=0.25\textwidth]{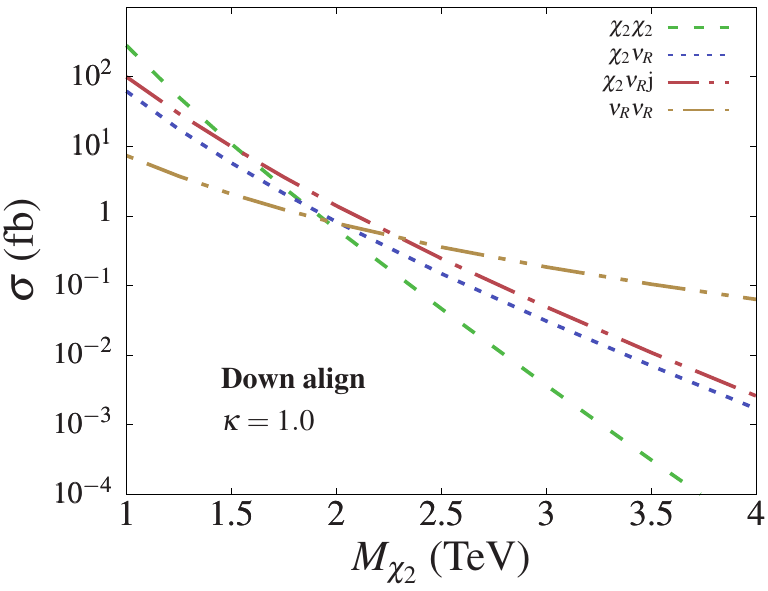}\label{fig:CSset304}}
\caption{Cross-sections of different production modes of sLQs [(a) -- (d)] and vLQs with $\kappa=0$ [(e) -- (h)] and $\kappa=1$ [(i) -- (l)]. We also show the cross-section of RHN pair production through $t$-channel LQ exchange. The LQ single productions and RHN pair production  process are computed for $\lambda (\Lambda)=1$.}
\label{fig:CS_sLQ_vLQ}
\end{figure*}

\section{LHC Phenomenology}\label{sec:lhc}
\noindent
We implement the phenomenological Lagrangian terms in {\sc FeynRules}~\cite{Alloul:2013bka}
to generate  {\sc Universal FeynRules Output} (UFO) files~\cite{Degrande:2011ua} needed for {\sc MadGraph5}~\cite{Alwall:2014hca} to generate the signal and background events at the leading order. We use the default dynamical scale choice in {\sc MadGraph5} for event generation. Whenever available, we account for higher-order cross-sections with appropriate $K$ factors. In particular, among the signal processes, we use a typical $K$-factor of $1.5$ for the pair production of the sLQs~\cite{Kramer:2004df, Mandal:2015lca,Borschensky:2022xsa}. We pass the parton-level events first through {\sc Pythia8}~\cite{Sjostrand:2014zea} for showering and hadronisation and then  through {\sc Delphes3}~\cite{deFavereau:2013fsa} for detector simulation with the default CMS card. The jets are reconstructed from the {\sc Delphes} tower objects with anti-$k_T$ clustering algorithm~\cite{Cacciari:2008gp} in {\sc FastJet}~\cite{Cacciari:2011ma}. We use jets of two different radii in our analysis: (a) AK4 jets with $R=0.4$ and (b) AK8 fatjets with $R=0.8$. 

\subsection{Production at the LHC}
\noindent
As mentioned earlier, LQs can be produced at the LHC as resonances in pairs (mainly through strong interactions) or singly (mediated by the LQ-quark-RHN coupling, $\lambda/ \Lambda$). Once produced, the LQs exclusively decay to RHN-jet pairs. A LQ can also appear in the $t$ channel  and contribute to the RHN pair production, $qq\to\n_R\n_R$. We show the cross-sections of different production processes for $M_{\n_R}=500$ GeV and $\lm=1$ (for sLQs) or $\Lm =1$ (for vLQs) with $\kp=0$ and $1$ (the cross-sections of the LQ-production processes except the $t$-channel one increase with the extra $g\chi\chi$ coupling) in Fig.~\ref{fig:CS_sLQ_vLQ}. The  $t$-channel  cross-sections are smaller than the pair and single LQ production cross-sections for lower LQ masses. Moreover, it is more suppressed for the sLQs than the vLQs. However, the $qq\to \nu_R\nu_R$ process becomes important for large couplings as its cross-section grows as $\lm^4$ or $\Lm^4$.\footnote{Though not significant for our study, the asymmetric pair production of the doublet LQs, where two different components are produced simultaneously, can also become non-negligible for very large couplings~\cite{Dorsner:2022ibm,Dorsner:2021chv}.} If LQs couple with the first-generation quarks, the RHN pair production cross-section gets an additional boost from the larger PDFs.

Since there is no direct experimental bound on the LQs decaying exclusively through RHNs, they can be even lighter than a TeV. In this paper, however, we mainly focus on the  $M_{\rm LQ} \geq 1$ TeV and $M_{\n_R}\sim 500$ GeV region of the parameter space. In our computations, we include the contributions of all the above $\n_R$-production processes to estimate the signal significance.

 The RHN pair production processes can be classified in terms of the number of charged leptons in the final state:
\begin{enumerate}
\item[(a)]  \textbf{Monolepton final state}: Ones with a muon accompanied by jet(s), fatjet(s) (from the hadronic decays of heavy bosons generated in the $\n_R$ decay), and some missing $E_T$. The different production modes contribute in the following manner:
\begin{align*}
    pp\ \to\ \left\{\begin{array}{l}
    \phi\phi/\chi\chi\\
    \phi/\chi\,\nu_{R}\ (+j)\\
    \nu_R\nu_{R}\ (+j)
    \end{array}\right\}\to& \left\{\begin{array}{l}
    (j\nu_{R})(j\nu_{R})\\
    (j\nu_{R})\nu_{R}\ (+j)\\
    \nu_R\nu_{R}\ (+j)\\
    \end{array}\right\}\nn\\\to&\ \m^{\pm}\,W^{\mp}_h\,Z_h \n_L+ \mbox{jet(s)}.
\end{align*}
\item[(b)] \textbf{Dilepton final state:} Ones with a opposite-sign muon pair ($\m^+\m^-$) plus jet(s) and fatjet(s). The different processes contribute to the dilepton final states as
\begin{align*}
    pp\ \to\ \left\{\begin{array}{l}
    \phi\phi/\chi\chi\\
    \phi/\chi\,\nu_{R}\ (+j)\\
    \nu_R\nu_{R}\ (+j)
    \end{array}\right\}\to& \left\{\begin{array}{l}
    (j\nu_{R})(j\nu_{R})\\
    (j\nu_{R})\nu_{R}\ (+j)\\
    \nu_R\nu_{R}\ (+j)
    \end{array}\right\}\nn\\
    \to&\left\{\begin{array}{l}
    \m^{\pm}\m^{\mp}\,W^{\pm}_h W^{\mp}_h+ \mbox{jet(s)}\\
    \nu_L\nu_L Z_h Z_\m + \mbox{jet(s)}
    \end{array}\right\}.
\end{align*}
\end{enumerate}
The subscripts $h$ and $\m$ denote the hadronic and leptonic decays of vector bosons, respectively. As shown above, a dimuon final state can arise in two ways. However, as one applies the $Z$-veto cut to suppress the large Drell-Yan dilepton background, the signal events from the leptonic decays of $Z$ are also cut out. In principle, one can also get more than two muons in the final state, where the extra leptons come from the decays of the vector bosons. However, since the leptonic branchings of the heavy vector bosons are smaller than their hadronic branchings, we do not consider final states with more than two muons.

In this paper, we follow a strategy of combining the contributions of multiple signal processes, which follows from our earlier papers~\cite{Mandal:2012rx, Mandal:2015vfa} where we showed how one could systematically combine pair and single production processes leading to the same final states without double counting. Later, in Refs.~\cite{Mandal:2016csb,Chandak:2019iwj,Bhaskar:2020gkk,Bhaskar:2021gsy}, we further demonstrated the usefulness of it. Here, we extend this strategy to the pair production of RHNs.

\begin{table}[t!]
\caption{Cross-sections of the major background processes without any cut. The higher-order cross-sections are taken from the literature; the corresponding QCD orders are shown in the last column. We use these cross-sections to compute the $K$ factors to incorporate the higher-order effects.}
\centering{\footnotesize\renewcommand\baselinestretch{1.25}\selectfont
\begin{tabular*}{\columnwidth}{l @{\extracolsep{\fill}} crc }
\hline
\multicolumn{2}{l}{Background } & $\sg$ & QCD\\ 
\multicolumn{2}{l}{processes}&(pb)&order\\\hline\hline
\multirow{2}{*}{$V +$ jets~ \cite{Catani:2009sm,Balossini:2009sa}  } & $Z +$ jets  &  $6.33 \times 10^4$& N$^2$LO \\ 
                & $W +$ jets  & $1.95 \times 10^5$& NLO \\ \hline
\multirow{3}{*}{$VV +$ jets~\cite{Campbell:2011bn}}   & $WW +$ jets  & $124.31$& NLO\\ 
                  & $WZ +$ jets  & $51.82$ & NLO\\ 
                   & $ZZ +$ jets  &  $17.72$ & NLO\\ \hline
\multirow{3}{*}{Single $t$~\cite{Kidonakis:2015nna}}  & $tW$  &  $83.10$ & N$^2$LO \\ 
                   & $tb$  & $248.00$ & N$^2$LO\\ 
                   & $tj$  &  $12.35$ & N$^2$LO\\  \hline
$tt$~\cite{Muselli:2015kba}  & $tt +$ jets  & $988.57$ & N$^3$LO\\ \hline
\multirow{2}{*}{$ttV$~\cite{Kulesza:2018tqz}} & $ttZ$  &  $1.05$ &NLO+N$^2$LL \\
                   & $ttW$  & $0.65$& NLO+N$^2$LL \\ \hline
\end{tabular*}}
\label{tab:Backgrounds}
\end{table}
\begin{table*}
\caption{The selection cuts applied on the monolepton and dilepton final states. The $b$ veto reduces the top backgrounds and fj denotes a fatjet.}
\centering{\footnotesize\renewcommand\baselinestretch{1.25}\selectfont
\begin{tabular*}{\textwidth}{l @{\extracolsep{\fill}}ll}
\hline \multirow{2}{*}{Selection cuts} & \multicolumn{2}{c}{Channel}\\\cline{2-3}
& Monolepton & Dilepton\\ \hline\hline
$\mathfrak{C}_1$: Selection of high $p_{\rm T}$ Leptons and jets &  \begin{tabular}[c]{@{}l@{}}$p_T(\m)>200$ GeV, \\$p_T(j_1)>200$ GeV,\\ No $b$-tagged jet \\\end{tabular} & \begin{tabular}[c]{@{}l@{}}$p_{\rm T}(\m)>220$ GeV,\\  $p{\rm _T}(j_1)>200$ GeV,\\No $b$-tagged jet 
~\end{tabular}\\ \hline
$\mathfrak{C}_2$: Identification of fatjet  & \begin{tabular}[c]{@{}l@{}}$p_{\rm T}({\rm fj})>80$ GeV,\\ $\tau_{21}<0.3$, \\ $65<M({\rm fj})<100$ GeV,\\ $\Delta R({\rm fj},\m)>1.0$\end{tabular} & \begin{tabular}[c]{@{}l@{}}$p_{\rm T}({\rm fj_{\phi(\chi)}})>120~(180)$ GeV,\\ $\tau_{21}<0.3$, \\ $65<M({\rm fj})<100$ GeV,\\ $\Delta R({\rm fj},\m)>0.8$ \\ \end{tabular} \\ \hline
$\mathfrak{C}_3$: Dilepton invariant mass & --- 
& $M(\m,\m)>150$ GeV\\
\hline
$\mathfrak{C}_4$: Scalar cuts &  $S_{\rm T}>1200$ GeV, $\slashed E_T > 150$ GeV, $\rm fj_{H_T}>600$ GeV &  $S_{\rm T}>1400$ GeV, $\rm fj_{H_T}>600$ GeV\\
\hline
\end{tabular*}}
\label{tab:cutsonsLQ}
\end{table*}

\begin{table*}
\caption{Cut flows for two $\phi_1$ benchmarks ($\lm\to\{0,1\}$) and the relevant background processes at luminosity $\mathcal{L}=3$ ab$^{-1}$.}
\centering
{\footnotesize\renewcommand\baselinestretch{1.25}\selectfont
\begin{tabular*}{\textwidth}{l @{\extracolsep{\fill}}rrrr}
\hline
\multirow{2}{*}{\bf $\blacksquare$~Monolepton final state}
& \multicolumn{2}{c}{Selection cuts }\\\cline{2-5} 
& $\mathfrak{C}_1$ & $\mathfrak{C}_2$ & $\mathfrak{C}_3$ & $\mathfrak{C}_4$
\tabularnewline
\hline\hline
\multicolumn{5}{l}{Signal benchmarks} \\
\hline
$M_{\phi_1}=1250$ GeV, $M_{\nu_{R}}=500$ GeV (pair production)         
& $270$           & $126$           & $126$           & $106$\\
$M_{\phi_1}=1250$ GeV, $M_{\nu_{R}}=500$ GeV (single production)       
& $291$           & $159$           & $159$           & $117$\\
\hline
 & \multicolumn{3}{r}{Total number of monolepton signal events:} & $223$\\
\hline
\multicolumn{5}{l}{Background processes} \\

\hline
$W_{\ell}~(+ 2j)$              

&$4.53\times10^7$ & $2.02\times10^6$        & $2.02\times10^6$         & $39346$ \tabularnewline

$W_{\ell} Z_{h} ~(+ 2j)$

& $2.55\times10^5$          & $74148$          & $74148$        & $3895$    \tabularnewline

$Z_{\ell}~(+ 2j)$

&$1.46\times10^7$ & $7.93\times10^5$        & $7.93\times10^5$         & $2640$ \tabularnewline

$ W_{\ell}W_{h}~(+ 2j)$

& $2.00\times10^5$          & $71229$          & $71229$        & $2483$    \tabularnewline
$t_h W_\ell+t_\ell W_h$

& $1.52\times10^5$          & $45436$          & $45436$        & $1729$  \tabularnewline

$t_{\ell} + b/j$                  

& $3313$          & $555$          & $555$        & $22$  \tabularnewline

\hline
 & \multicolumn{3}{r}{Total number of background events:} & $50115$\\
\hline
\multirow{2}{*}{\bf $\blacksquare$~Dilepton final state}& \multicolumn{2}{c}{Selection cuts}\\\cline{2-5} 
& $\mathfrak{C}_1$ & $\mathfrak{C}_2$ & $\mathfrak{C}_3$ & $\mathfrak{C}_4$
\tabularnewline
\hline\hline
\multicolumn{5}{l}{Signal benchmarks} \\
\hline
$M_{\phi_1}=1250$ GeV, $M_{\nu_{R}}=500$ GeV (pair production)       
& $339$           & $166$           & $119$           & $118$\\
$M_{\phi_1}=1250$ GeV, $M_{\nu_{R}}=500$ GeV (single production)      
& $194$           & $113$           & $73$           & $70$\\
\hline
 & \multicolumn{3}{r}{Total number of dilepton signal events:} & $188$\\
\hline
\multicolumn{5}{l}{Background processes} \\

\hline
$W_{\ell} Z_{\ell} ~(+ 2j)$                  

& $39685$          & $2125$          & $632$        & $101$    \tabularnewline

$ W_{\ell}W_{\ell}~(+ 2j)$ 

& $31093$          & $1422$          & $514$        & $51$    \tabularnewline

$t_{\ell}W_{\ell}$       

& $19694$          & $1062$          & $218$        & $28$  \tabularnewline  

$t_{\ell} t_{\ell} ~(+ 2j)$             

& $21256$           & $1097$           & $60$        & $20$    \tabularnewline

\hline
 & \multicolumn{3}{r}{Total number of background events:} & $200$\\
\hline
\end{tabular*}}
\label{tab:cut-flow}
\end{table*}

\begin{figure*}
\centering
\captionsetup[subfigure]{labelformat=empty}
\subfloat[\quad\quad(a)]{\includegraphics[width=0.25\textwidth]{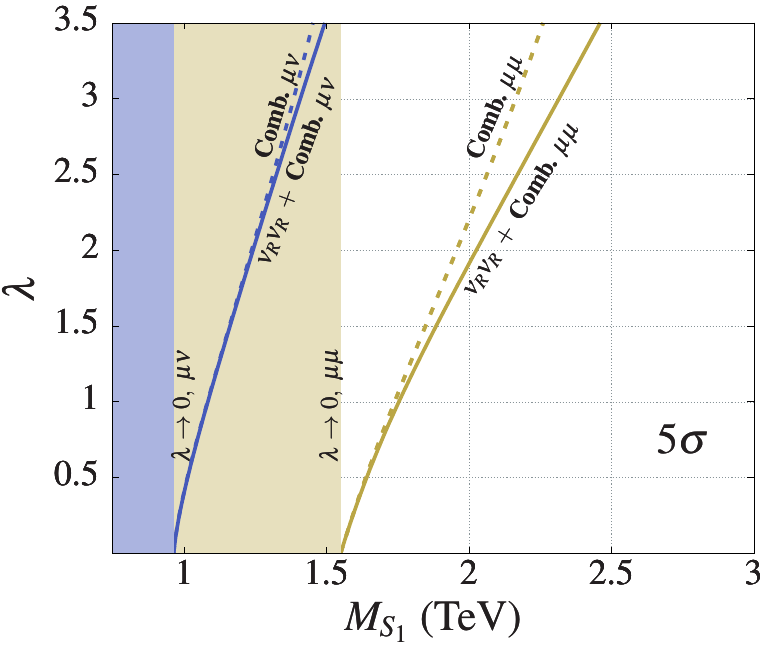}\label{fig:mslq1_lam5}}\hfill
\subfloat[\quad\quad(b)]{\includegraphics[width=0.25\textwidth]{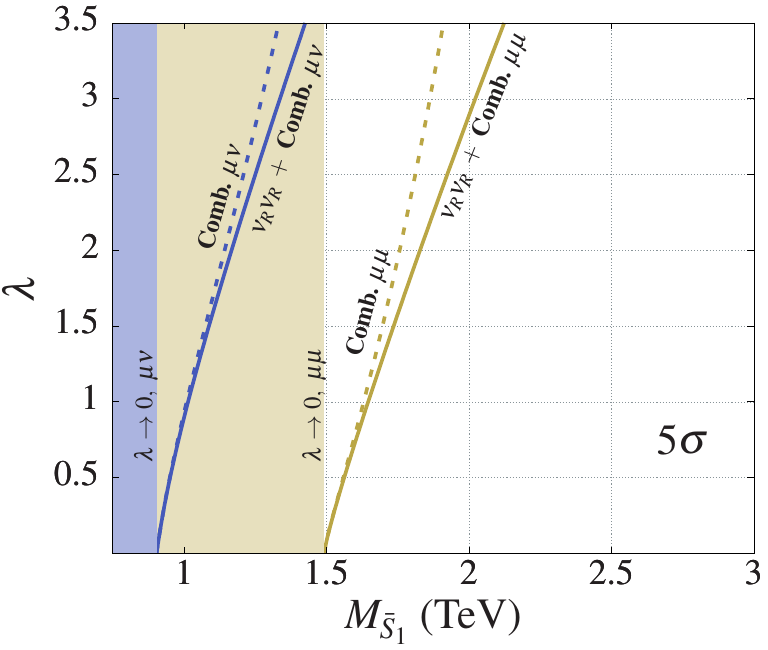}\label{fig:mslq2_lam5}}\hfill
\subfloat[\quad\quad(c)]{\includegraphics[width=0.25\textwidth]{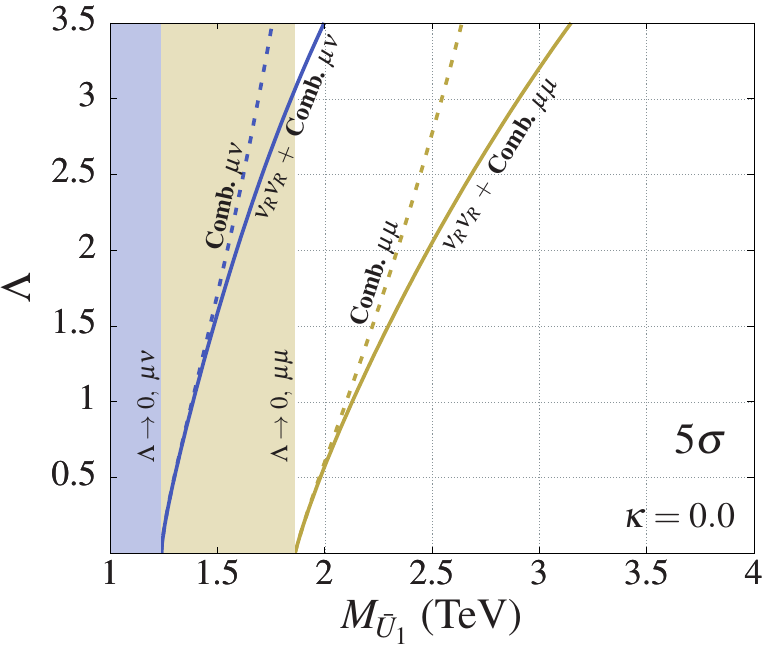}\label{fig:mvlq1k0_lam5}}\hfill
\subfloat[\quad\quad(d)]{\includegraphics[width=0.25\textwidth]{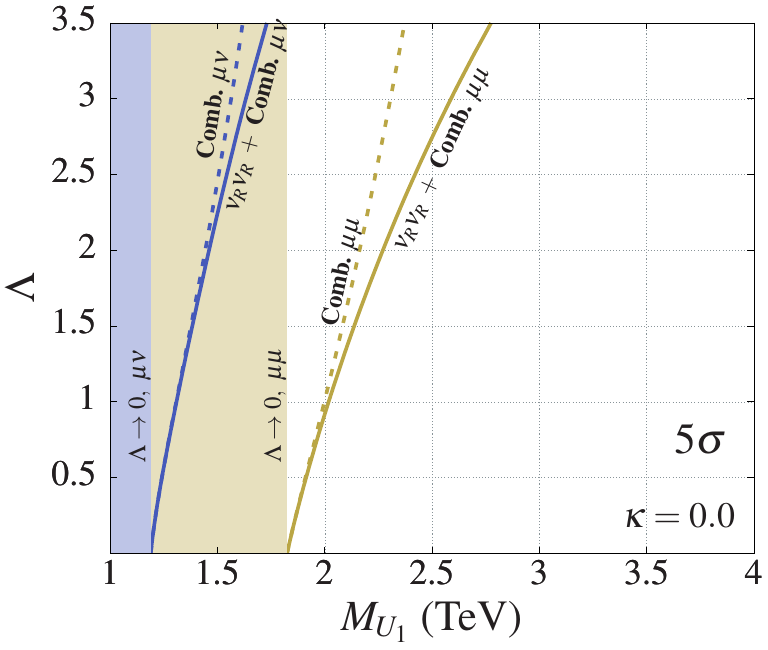}\label{fig:mvlq2k0_lam5}}\\
\subfloat[\quad\quad(e)]{\includegraphics[width=0.25\textwidth]{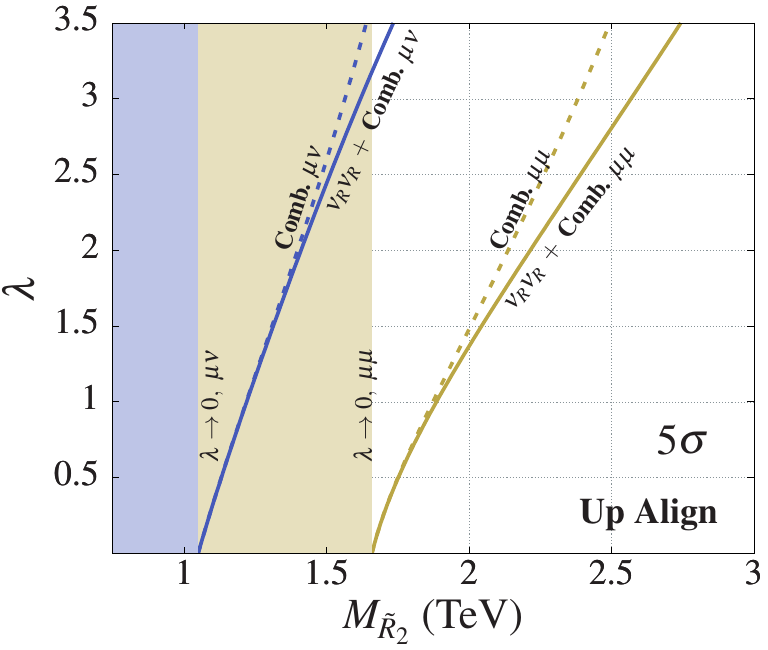}\label{fig:mvlq2k0_lam2_mixing}}\hfill
\subfloat[\quad\quad(f)]{\includegraphics[width=0.25\textwidth]{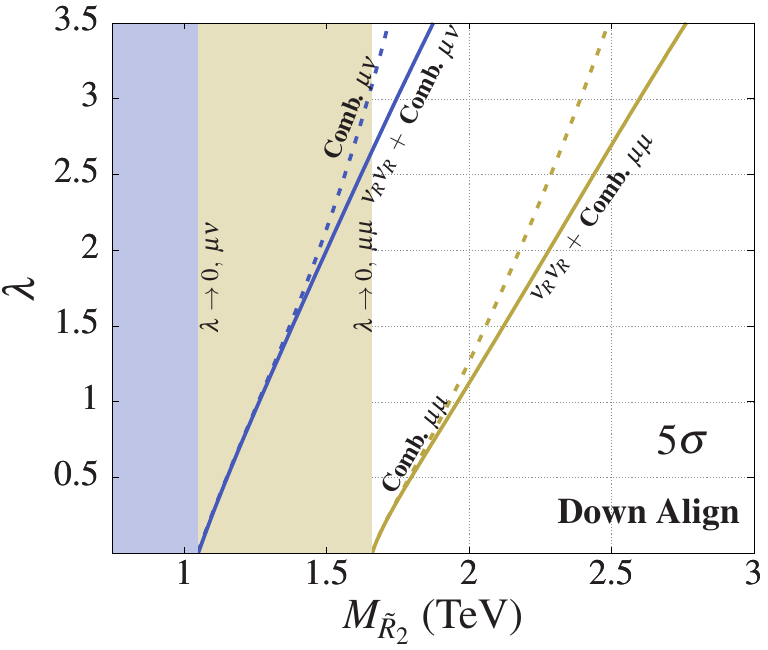}\label{fig:mslq2_lam2_mixing}}\hfill
\subfloat[\quad\quad(g)]{\includegraphics[width=0.25\textwidth]{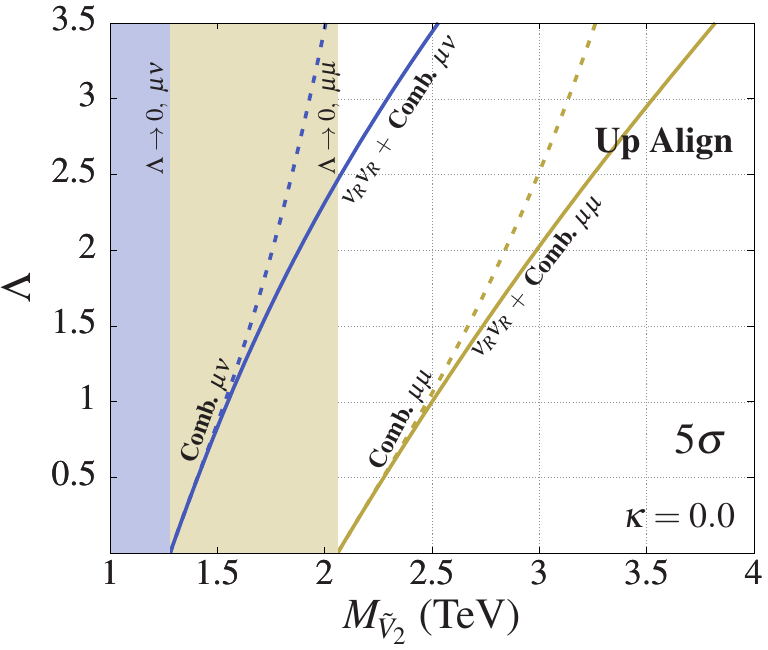}\label{fig:mvlq2k0_lam5_mixing}}\hfill
\subfloat[\quad\quad(h)]{\includegraphics[width=0.25\textwidth]{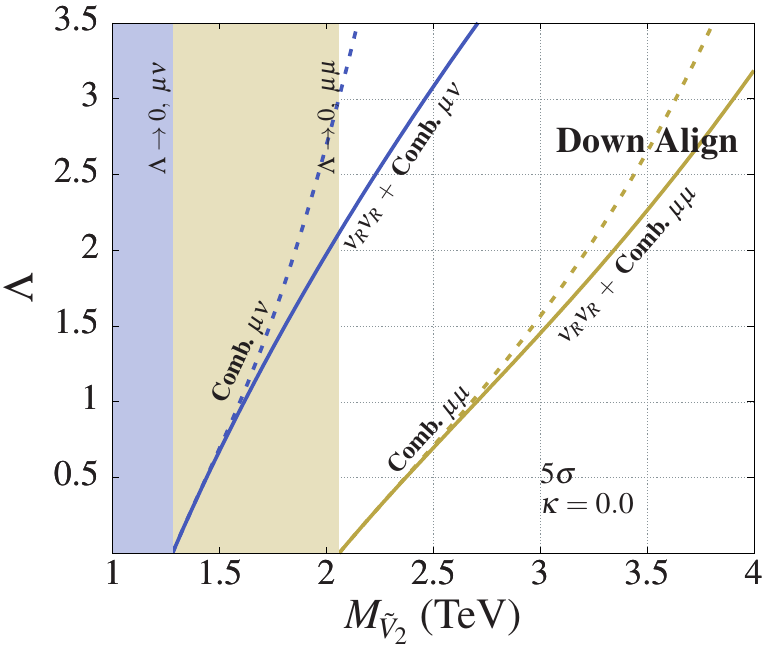}\label{fig:mslq2_lam5_mixing}}
\\
\subfloat[\quad\quad(i)]{\includegraphics[width=0.25\textwidth]{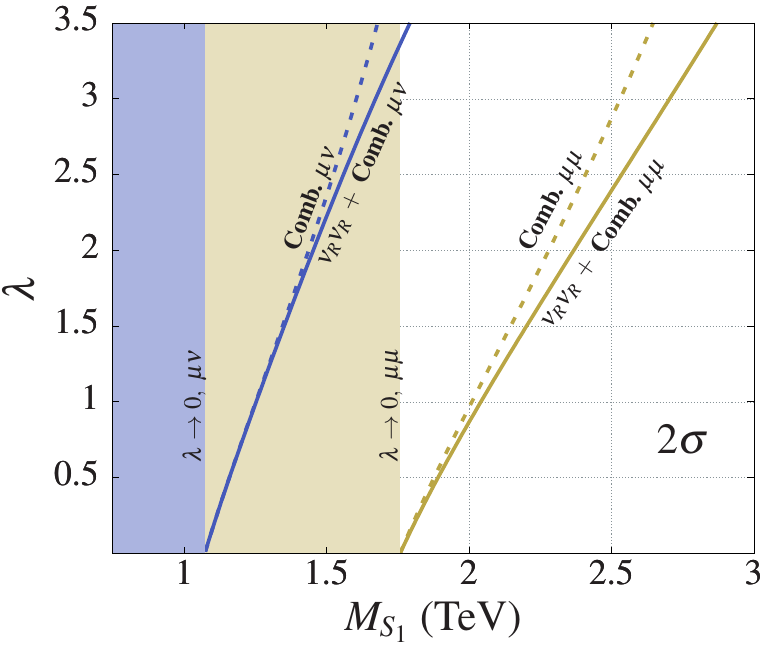}\label{fig:mslq1_lam2}}\hfill
\subfloat[\quad\quad(j)]{\includegraphics[width=0.25\textwidth]{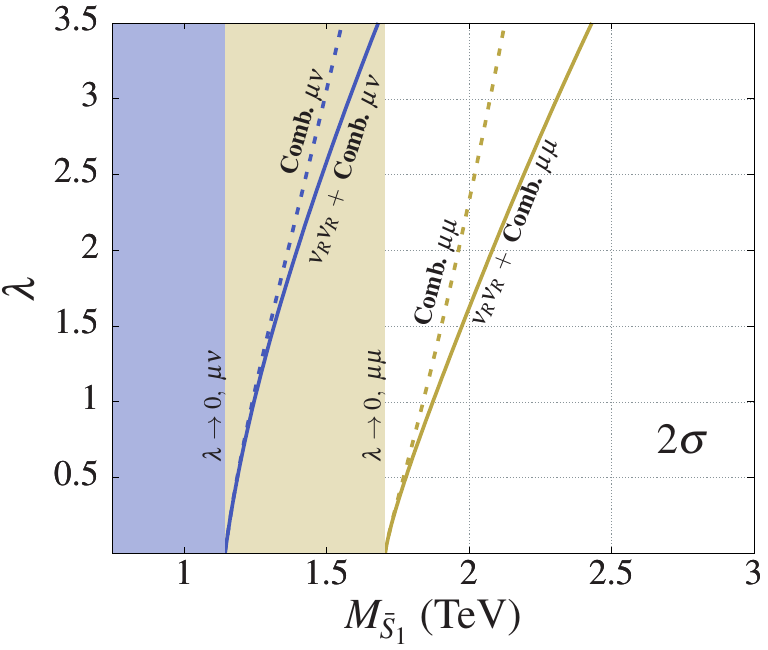}\label{fig:mslq2_lam2}}\hfill
\subfloat[\quad\quad(k)]{\includegraphics[width=0.25\textwidth]{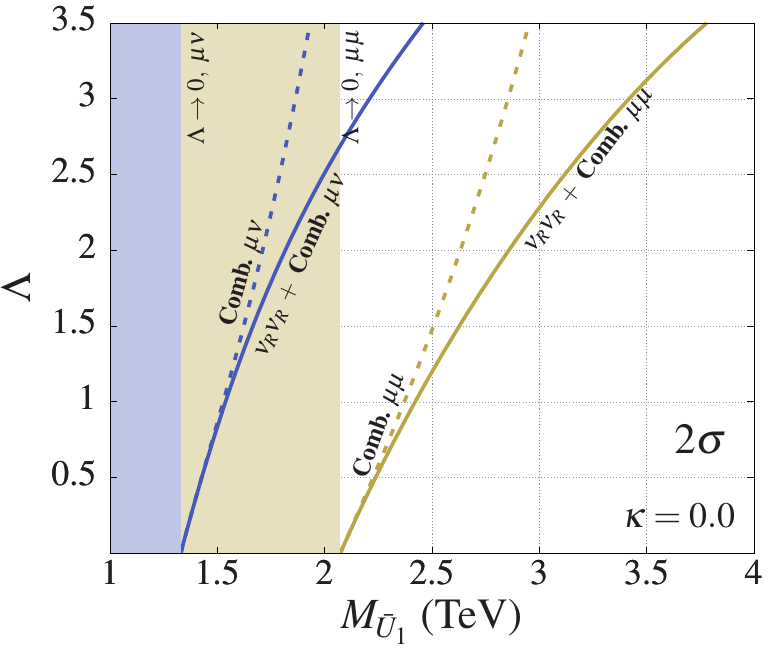}\label{fig:mvlq1k0_lam2}}\hfill
\subfloat[\quad\quad(l)]{\includegraphics[width=0.25\textwidth]{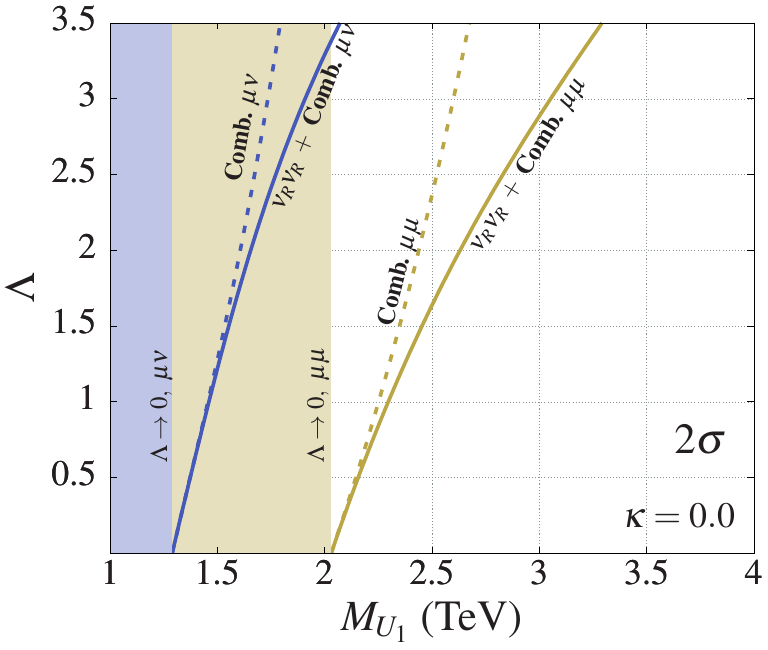}\label{fig:mvlq2k0_lam2}}\\
\subfloat[\quad\quad(m)]{\includegraphics[width=0.25\textwidth]{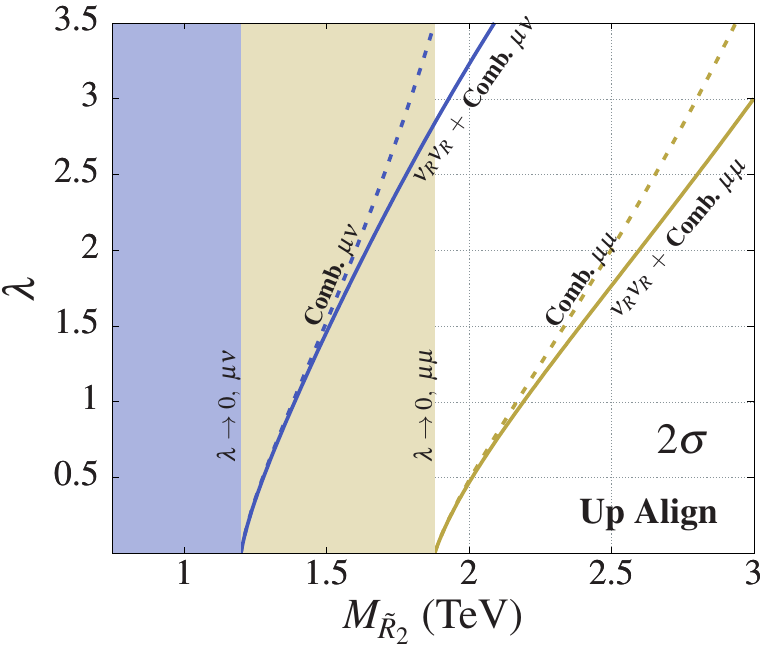}\label{fig:mvlq1k0_lam2_mixing}}\hfill
\subfloat[\quad\quad(n)]{\includegraphics[width=0.25\textwidth]{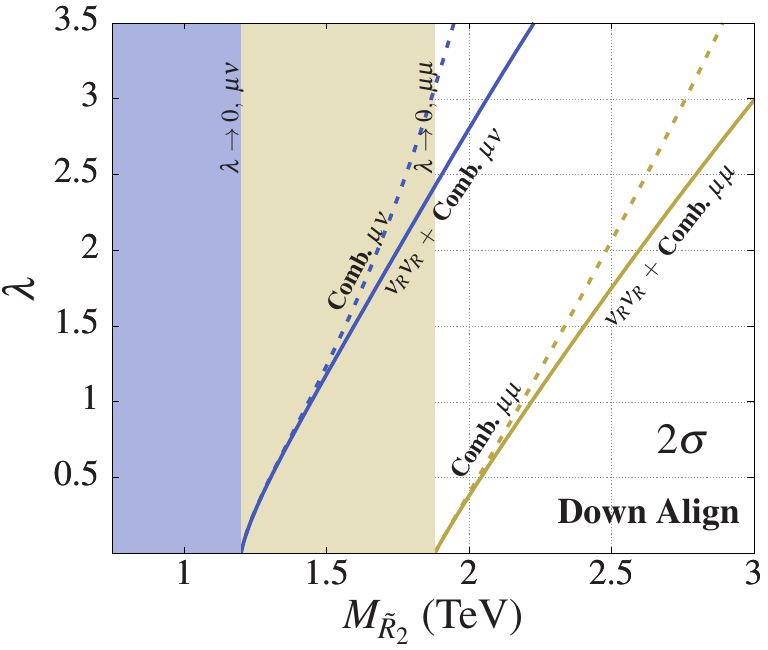}\label{fig:mslq1_lam2_mixing}}\hfill
\subfloat[\quad\quad(o)]{\includegraphics[width=0.25\textwidth]{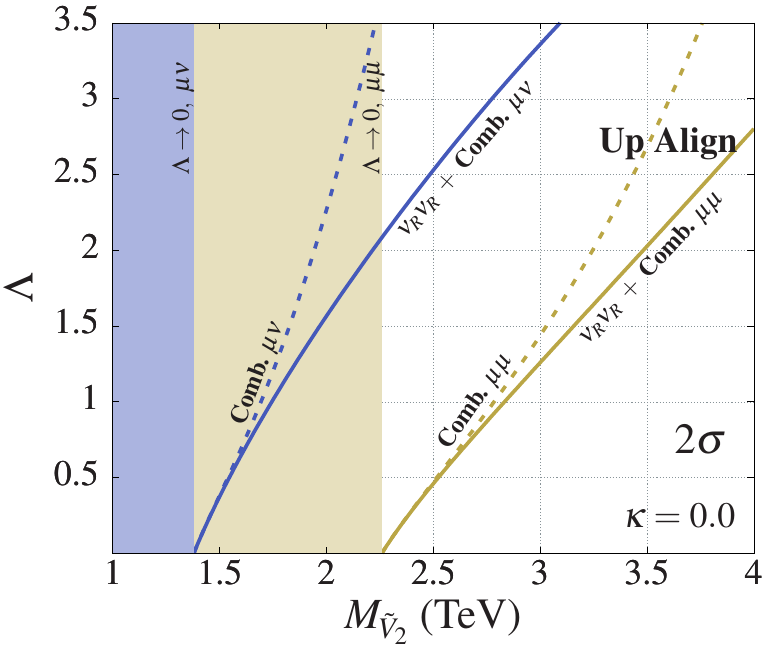}\label{fig:mvlq1k0_lam5_mixing}}\hfill
\subfloat[\quad\quad(p)]{\includegraphics[width=0.25\textwidth]{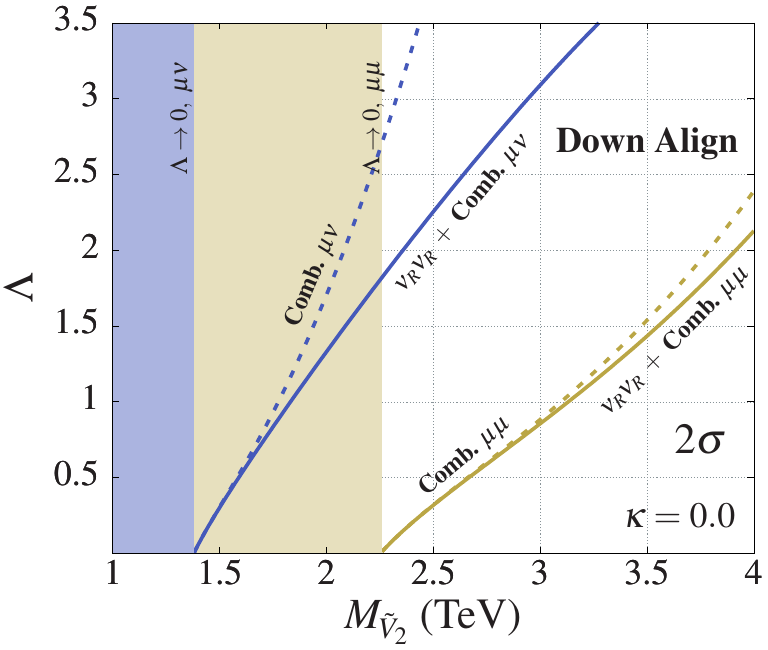}\label{fig:mslq1_lam5_mixing}}
\caption{The least values of the new coupling ($\lm$ or $\Lm$) needed to observe the signals with $5\sigma$ (top two rows) and $2\sigma$ (bottom two rows) significances as functions of masses at the HL-LHC. These plots are generated for $M_{\n_R}=500$ GeV and $\kp=0$ for the vLQS. The QCD regions ($\lm$, $\Lm\to 0$; dominated by LQ pair productions) in the monolepton and dilepton channels are shown with solid colours; the dashed lines are obtained by combining the LQ pair and single production events. Combining single-production events with pair-production events enhances the prospects. However, the prospects improve even further for high couplings since the RHN pair production via LQ exchange also contributes to the signals and enhances the significances (solid lines).}
\label{fig:MLQ_lam_25}
\end{figure*}
\begin{figure*}
\centering
\captionsetup[subfigure]{labelformat=empty}
\subfloat[\quad\quad(a)]{\includegraphics[width=0.25\textwidth]{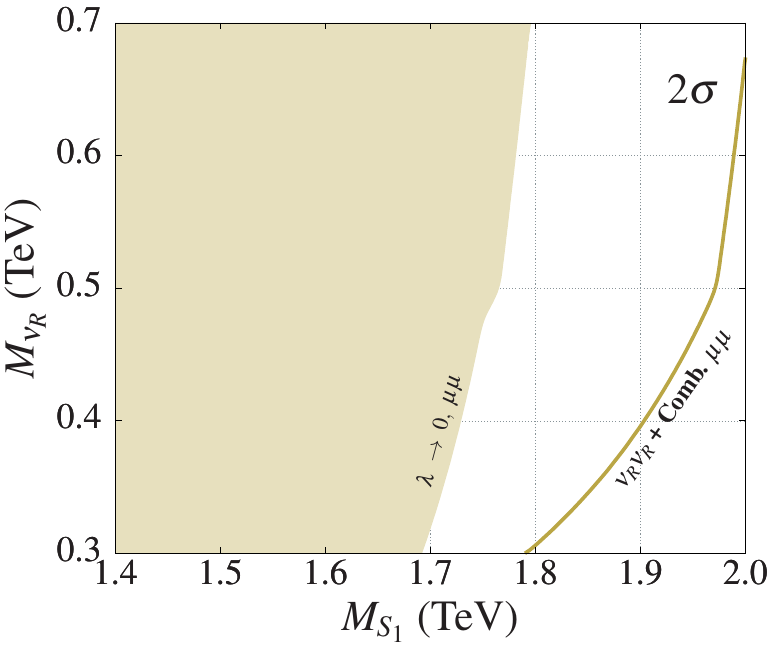}\label{fig:Mslq_mnur_2}}\hfill
\subfloat[\quad\quad(b)]{\includegraphics[width=0.25\textwidth]{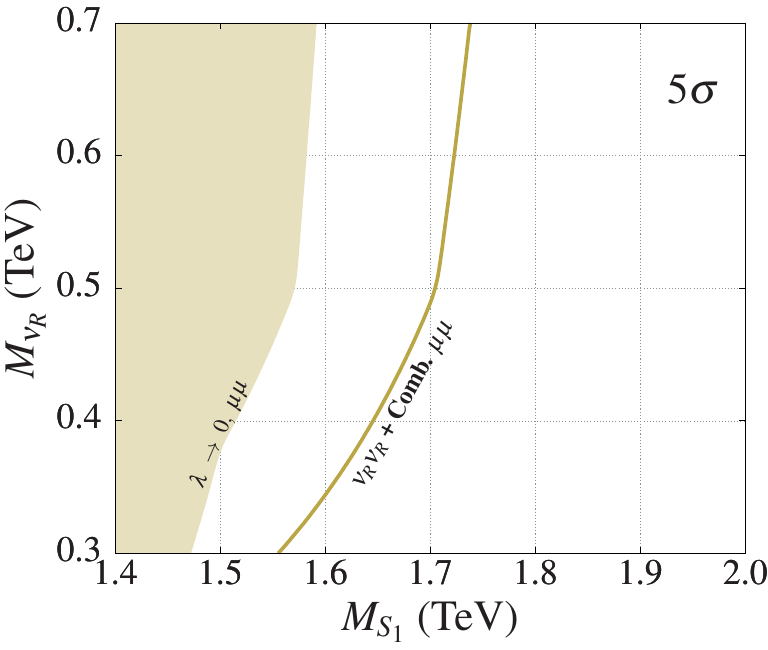}\label{fig:MSLQ_Mnur_5}}\hfill
\subfloat[\quad\quad(c)]{\includegraphics[width=0.25\textwidth]{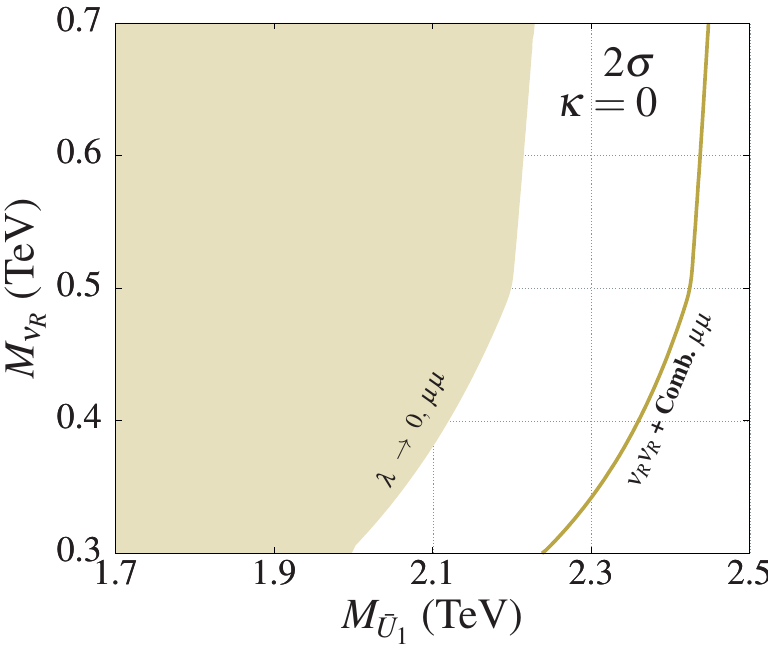}\label{fig:MVLQ_Mnur_2}}\hfill
\subfloat[\quad\quad(d)]{\includegraphics[width=0.25\textwidth]{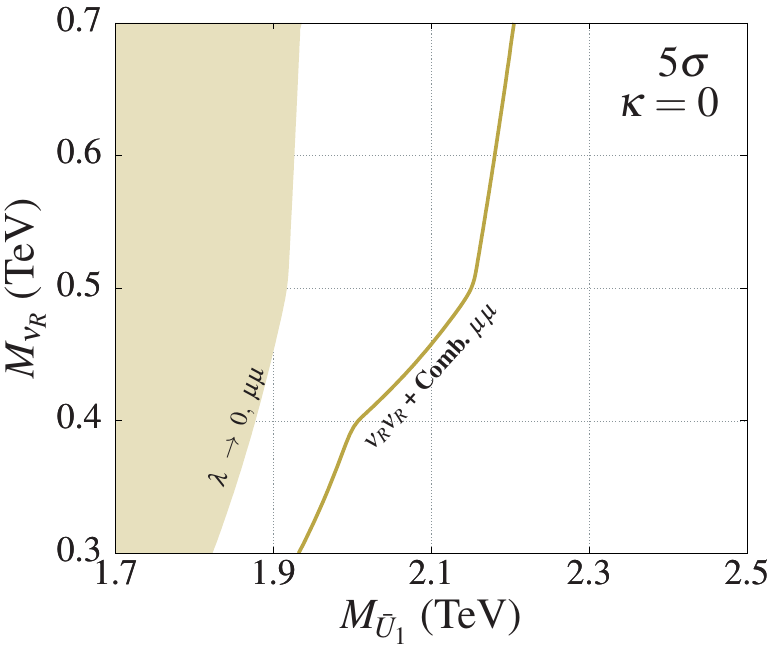}\label{fig:MVLQ_Mnur_5}}
\caption{The [(a), (c)] $2\sigma$ and [(b), (d)] $5\sigma$ contours in the dilepton mode on the $M_{S_1}/M_{\bar U_1}$--$M_{\nu_R}$ plane. The combined contours are obtained for $\lm$, $\Lm=1$. }
\label{fig:MSLQ_Munr}
\end{figure*}

\subsection{Signals, cuts and the background}
\noindent We define our inclusive signal selection criteria for monolepton and dilepton events as: 
\begin{enumerate}
\item[(a)]
A monolepton event must have
\begin{itemize}[leftmargin=-.1cm]
    \item [--] exactly one high-$p_T$ muon and
    \item [--] at least one high-$p_T$ AK4 jet and at least one AK8 fatjet.  
\end{itemize}
\item[(b)]
A dilepton event must have
\begin{itemize}[leftmargin=-.1cm]
    \item [--] a pair of opposite-sign muons, at least one of which should have high-$p_T$.
    \item [--] at least one high-$p_T$ AK4 jet and at least one AK8 fatjet. 
\end{itemize}
\end{enumerate}

The relevant background processes are listed in Table~\ref{tab:Backgrounds}. While generating the processes with very high cross-sections, we apply some basic generational-level cuts to save computation time. For the dilepton final state, the $W_\ell+ $jets process can act as a background since a jet can be misidentified as a lepton. In fact, it is one of the major backgrounds for the RHN searches with same-sign dilepton final states. Since we consider only opposite-sign lepton pairs, in our case, its contribution is not important as the jet-faking-lepton efficiency is very small, $\sim 10^{-4}$~\cite{Curtin:2013zua}.

We show the cuts we apply on the two signals and the relevant backgrounds in Table~\ref{tab:cutsonsLQ}. In Table~\ref{tab:cut-flow}, we show how these cuts affect various background processes. As examples, we also show the effect of the cuts on the signals for two $\phi_1$ benchmarks.

\section{HL-LHC Prospects}
\label{sec:HL-LHC}
\noindent
We show the HL-LHC prospects of the mono- and dilepton channels for $M_{\n_R}=500$ GeV in Fig.~\ref{fig:MLQ_lam_25} with 
the $5\sigma$ (discovery) and $2\sigma$ ($\sim$ exclusion) significance contours on the $M_{\rm LQ}$--$\lm/\Lm$ plane for each LQ mediator.  We have used the simple models introduced in Sec.~\ref{subsec:simplifiedmodels} to estimate the significances at $3$ ab$^{-1}$; see Appendix~\ref{sec:appendixA} for our method of estimating the signal significance, $\mathcal{Z}$. (As explained, the mapping of the simple models to the actual models is straightforward for the singlet LQ models. For each doublet LQ, there are two possibilities depending on whether it is aligned with the up- or down-type quarks. However, from the cases covered in Fig.~\ref{fig:CS_sLQ_vLQ}, we can map the simple models to the components of doublet LQs.) From the plots in Fig.~\ref{fig:MLQ_lam_25}, we see that, generally, the dilepton channel has better prospects than the monolepton one. This is because the branching ratio (BR) of the RHN in the $W\m$ mode is roughly twice that of the $Z\n$ mode, i.e., BR($\n_R\to W^\pm\m^\mp$)$\approx 2$ BR($\n_R\to Z\n$), and the dilepton cuts lead to better yields due to the high $\m$-detection efficiency.

For the weak-singlet LQs and small values of $\lm$ or $\Lm$ (small to suppress the LQ single productions and the $t$-channel LQ exchange but not enough to form displaced vertices), the $5\sigma$ discovery reaches for $S_1$ and $\bar S_1$ in the pair production mode in the dilepton channel can go as high as $1.6$ and $1.5$ TeV, respectively. But, for $\lm=1$, the reaches enhance to $1.8$ and $1.6$ TeV once the single production and the $t$-channel RHN-pair production contributions are combined in the signal. 
For $\bar U_1$ and $U_1$, the $5\sigma$ reaches in the pair-only mode are about $1.9$ and $1.8$ TeV, respectively, for the same RHN mass and $\kp=0$, which increase to $2.1$ and $2.0$ TeV when the single production events are combined in the signal. For $\lm=1$,  the actual reaches are slightly better as the $qq\to\n_R\n_R$ process also contributes. As expected, the $t$-channel process becomes important only for large couplings.

In the absence of discovery, the $2\sigma$ values provide rough estimates of the exclusion limits. For $M_{\n_R}=500$ GeV, the HL-LHC can exclude up to $2.0$ and $1.9$ TeV in the cases of $S_1$ and $\bar S_1$, respectively. The exclusion limits for $\bar U_1$ and $U_1$ for $\kappa=0$ are $2.4$ and $2.3$ TeV, respectively. 

The vLQ numbers shown in Fig.~\ref{fig:MLQ_lam_25} should be considered conservative as we have put the extra gauge coupling to zero. Their prospects improve if this coupling is nonzero. For example, if $\kp=1$, the $5\sigma$ reaches for $\bar U_1$ and $U_1$ go to about $2.2$ and $2.1$ TeV, respectively, in the pair-only mode and to about $2.4$ and $2.2$ TeV in the combined mode (for $\Lm=1$). Similarly, the $2\sigma$ limits change to $2.6$ and $2.5$ TeV, respectively (for $\Lm=1$). The enhancement in each mode is proportional to the corresponding improvement in the cross-section (see Fig.~\ref{fig:CS_sLQ_vLQ}).
\begin{figure*}
\centering
\captionsetup[subfigure]{labelformat=empty}
\subfloat[\quad\quad\quad(a)]{\includegraphics[width=0.45\textwidth]{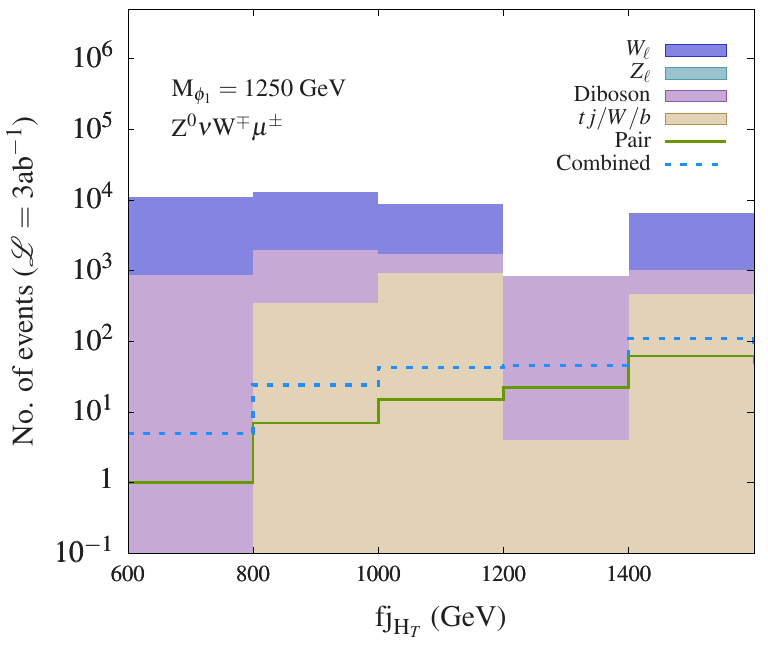}\label{fig:histo_ML}}\hspace{1cm}\subfloat[\quad\quad(b)]{\includegraphics[width=0.45\textwidth]{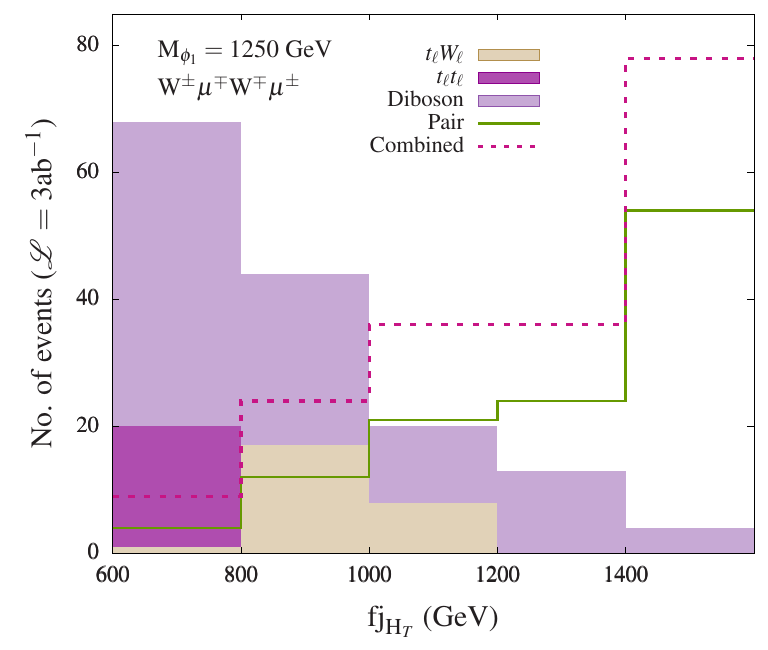}\label{fig:histo_DL}}
\caption{The distribution of signal and background events in (a) the monolepton channel and (b) the dilepton channel. The events are first  passed through $\mathfrak C_1$--$\mathfrak C_4$ (Table~\ref{tab:cutsonsLQ}) and then binned according to fj$_{\rm H_T}$ (the scalar sum of the transverse momenta of the fatjets). The fifth bin is open on the higher side, i.e., fj$_{\rm H_T}^{(5)}>1400$ GeV. The total number of events for every process is reported in the last row of Table~\ref{tab:cut-flow}.}
\label{fig:histograms}
\end{figure*}

The results are more promising for the doublet LQ models for two reasons. First, our selection criteria are insensitive towards the charge of the LQ. As a result, the contributions of the individual components to the signal add up. Second, one component of the doublets (either charge $1/3$ or $2/3$, depending on how the LQ is aligned) couples with a first-generation quark through the CKM mixing, which boosts the cross-sections. For $\lm=1$, the discovery reaches for both up-aligned $\tilde R_2$ (one with flavour off-diagonal couplings to the down-type quarks) and down-aligned $\tilde R_2$ (one with flavour off-diagonal couplings to the up-type quarks) go up to about $1.9$ TeV in the dilepton channel. Similarly, for $\Lm=1$ and $\kp=0$, the discovery reaches for up-aligned and down-aligned $\tilde V_2$ go up to $2.5$ and $3.0$ TeV, respectively. As expected, in the high coupling regions, the signal enhancements from the $qq\to\n_R\n_R$ process are more pronounced in the doublet cases than the singlet ones due to the first-generation quark in the initial states. Similar trends are observed in the doublet-LQ exclusion plots as well.

We draw the $2\sigma$ and $5\sigma$ contours on the $M_{S_1,\bar U_1}$--$M_{\n_R}$ planes in Fig.~\ref{fig:MSLQ_Munr} to demonstrate how the signal significance varies with $M_{\n_R}$. Since the RHNs appear only in the decay, we do not expect any significant dependence on $M_{\n_R}$ (we only show the $S_1,\bar U_1$ plots for illustration). We observe a drop in the sensitivity in the $M_{\nu_R} \ll M_{LQ}$ region. If the $M_{LQ}-M_{\nu_R}$ mass gap is large, the RHN becomes highly boosted, and the decay products of RHN become very collimated, making it difficult to isolate the $W$-like fatjet from the selected muon. This requires a different strategy, as discussed in Ref.~\cite{Mitra:2016kov}. 

\section{Summary and Conclusions}
\label{sec:conclu}
\noindent
In this paper, we analysed the LQ-mediated pair production of  RHNs at the LHC. This channel has received little attention in the literature and is yet to be explored experimentally. This is an interesting signal to probe if RHNs are within the TeV scale, a possibility realisable with the inverse seesaw mechanism for a wide range of neutrino parameters. In this mechanism, the RHNs are pseudo-Dirac type and are not restricted by the stringent same-sign dilepton-search bounds. Moreover, no direct experimental limits exist on LQs that exclusively couple with RHNs.

LQs can contribute to the RHN pair productions in two ways: 1) the RHNs come from the decay of LQs if they are lighter than the LQs, and 2) a $t$-channel LQ exchange can produce a pair of RHNs in the quark-fusion process. Since the LQs can be resonantly produced either in pairs or singly, we have three processes to produce RHN pairs. These three processes become significant in different regions of the parameter space. The mostly QCD-mediated LQ pair productions are dominant if the LQs are not very heavy and the new couplings $\lm, \Lm$ are not very large. The $t$-channel processes are significant if $\lm, \Lm$ are large ($\gtrsim 2$) and the exchanged LQs are very heavy (the resonant productions are beyond the reach of LHC). LQ single productions are important in the intermediate mass range (single productions of heavy particles are less phase-space-suppressed than the pair production) if $\lm, \Lm\gtrsim \mc O(1)$. 

Extending a strategy we proposed earlier, we defined the signals so that all three processes contribute to the signals, thus proving a complete coverage of the parameter space within the reach of the HL-LHC. However, for a conservative estimation of the HL-LHC prospects, we focused only on second-generation interactions in this paper. Our signals have either an opposite-sign muon pair or a single muon along with hadronically decaying bosons. We analysed the prospects of these channels at the HL-LHC and obtained the $5\sigma $ and $2\sigma$ significance contours. 

We found that LQ-mediated RHN production has excellent prospects at the HL-LHC. For scalar LQs, the discovery reaches vary from about $1$ TeV to close to $3$ TeV. For vector LQs, they are between $2$ TeV to $4$ TeV. Our analysis is generic and comprehensive as 1) we do not assume any specific property of the RHN except that it is lighter than a TeV and is pseudo-Dirac type (because of ISM) decaying to either a $W$ boson and a muon or a $Z/H$ boson and a neutrino, and 2) we considered all possible scalar and vector LQs that can exclusively couple to the second-generation RHN.

\section*{Model Files}
\noindent 
The UFO model files are available at \url{https://github.com/rsrchtsm/LQ\_RHN}.

\appendix
\section{Estimating $\mc Z$ score}
\label{sec:appendixA}
\noindent 
We showed how one could make use of the distribution of the data (rather than just the total number of events) to estimate the signal significance in Ref.~\cite{Bhaskar:2021gsy}. We follow the method outlined there.  After passing the events through the cuts in Table~\ref{tab:cutsonsLQ}, we first bin the data by the scalar sum of the transverse momenta of the fatjets,  ${\rm fj_{H_T}}$ (see Fig.~\ref{fig:histograms}). We then estimate the combined signal significance by Liptak-Stouffer (weighted) $\mc Z$ score~\cite{Stouffer1949,liptak1958combination}: 
\begin{equation}
    \mathcal{Z}=\frac{\sum_{i=1}^{5}w_i \mc Z_i}{\sqrt{\sum_{i=1}^{5}w_i^2}}.
\end{equation} 
Here, $\mc Z_i$ is the signal significance score in the $i^{\rm th}$ bin ($i\in\{1, 2, 3, 4,5\}$) computed as~\cite{Cowan:2010js}
\begin{align}
\mc Z_i = \sqrt{2\lt(N^i_S+N^i_B\rt)\ln\lt(\frac{N^i_S+N^i_B}{N^i_B}\rt)-2N^i_S}\,,
\label{eq:sig}
\end{align}
where $N^i_S$ and $N^i_B$ are the numbers of signal and background events in the bin. The corresponding weight is denoted as $w_i$. The weight is commonly taken as the inverse of the variance~\cite{Whitlock}. We set $w_i$ equal to the inverse of the square of the total error in the background events in the $i^{\rm th}$ bin, i.e., $w_i^{-1} = ({\rm statistical~error})^2=N_B^i$. 

\bibliography{bibliography.bib}
\bibliographystyle{JHEPCust}

\end{document}